\documentclass{aa}  

\usepackage{graphicx}
\usepackage{txfonts}
\usepackage{color}
\usepackage[colorlinks=true, allcolors=blue]{hyperref}
\usepackage{placeins}
\usepackage{multirow}

\usepackage{natbib}
\bibpunct{(}{)}{;}{a}{}{,}

\newcommand{\ergcm}[1]{$\times 10^{#1}$ erg cm$^{-2}$ s$^{-1}$\xspace}
\newcommand{\oergcm}[1]{$10^{#1}$ erg cm$^{-2}$ s$^{-1}$\xspace}
\newcommand{\uergcm}{erg cm$^{-2}$ s$^{-1}$\xspace}
\newcommand{\ergs}[1]{$\times 10^{#1}$ erg s$^{-1}$\xspace}

\newcommand{\uergs}{erg s$^{-1}$\xspace}
\newcommand{\hcm}[1]{$\times 10^{#1}$ cm$^{-2}$\xspace}
\newcommand{\ohcm}[1]{$10^{#1}$ cm$^{-2}$\xspace}
\newcommand{\uhcm}{cm$^{-2}$\xspace}
\newcommand{\expo}[1]{$\times 10^{#1}$\xspace}
\newcommand{\oexpo}[1]{$10^{#1}$\xspace}

\newcommand{\nhlmc}{N$_{\rm H}^{\rm LMC}$\xspace}
\newcommand{\nhGal}{N$_{\rm H}^{\rm Gal}$\xspace}

\newcommand{\Halpha}{H${\alpha}$\xspace}
\newcommand{\Hbeta}{H${\beta}$\xspace}
\newcommand{\ltsima}{$\buildrel < \over \sim$}
\newcommand{\lsim}{\lower.5ex\hbox{\ltsima}}
\newcommand{\gtsima}{$\buildrel > \over \sim$}
\newcommand{\gsim}{\lower.5ex\hbox{\gtsima}}

\newcommand{\xspec}{{\tt XSPEC}\xspace}

\newcommand{\arfgen}{{\tt arfgen}\xspace}
\newcommand{\rmfgen}{{\tt rmfgen}\xspace}

\newcommand{\eSASS}{{\tt eSASS}\xspace}
\newcommand{\nrta}{{\tt NRTA}\xspace}
\newcommand{\pattern}{{\tt PATTERN}\xspace}
\newcommand{\flag}{{\tt FLAG}\xspace}

\newcommand{\swift}{{\it Swift}\xspace}
\newcommand{\xmm}{{\it XMM-Newton}\xspace}
\newcommand{\nustar}{{\it NuSTAR}\xspace}
\newcommand{\ein}{{\it Einstein}\xspace}
\newcommand{\ROSAT}{{\it ROSAT}\xspace}

\newcommand{\srg}{{\it SRG}\xspace}
\newcommand{\ero}{\mbox{eROSITA}\xspace}
\newcommand{\asrc}{\mbox{eRASSU\,J050810.4$-$660653}\xspace}
\newcommand{\bsrc}{\mbox{RX\,J0501.6$-$7034}\xspace}
\newcommand{\csrc}{\mbox{eRASSt\,J044811.1$-$691318}\xspace}
\newcommand{\sasrc}{\mbox{J0508}\xspace}
\newcommand{\sbsrc}{\mbox{J0501}\xspace}
\newcommand{\scsrc}{\mbox{J0448}\xspace}

\begin{document} 

\title{SRG/eROSITA-triggered \xmm observations of three Be/X-ray binaries in the LMC: Discovery of X-ray pulsations}

\author{F. Haberl\inst{\ref{mpe}} \and
        C. Maitra\inst{\ref{mpe}} \and
        D. Kaltenbrunner\inst{\ref{mpe}} \and
        D.A.H. Buckley\inst{\ref{salt},\ref{uct}} \and
        I.M. Monageng\inst{\ref{salt},\ref{uct}} \and
        A.\,Udalski\inst{\ref{uw}} \and
        V. Doroshenko\inst{\ref{iaat}} \and
        L. Ducci\inst{\ref{iaat}} \and
        I. Kreykenbohm\inst{\ref{uerl}} \and
        P. Maggi\inst{\ref{oas}} \and
        A. Rau\inst{\ref{mpe}} \and
        G. Vasilopoulos\inst{\ref{oas}} \and
        P. Weber\inst{\ref{uerl}} \and
        J. Wilms\inst{\ref{uerl}}
       } 

\titlerunning{New Be/X-ray binary pulsars in the LMC}
\authorrunning{Haberl et al.}

\institute{
Max-Planck-Institut f{\"u}r extraterrestrische Physik, Gie{\ss}enbachstra{\ss}e 1, 85748 Garching, Germany\label{mpe}, \email{fwh@mpe.mpg.de}
\and
South African Astronomical Observatory, PO Box 9, Observatory Rd, Observatory 7935, South Africa\label{salt}
\and
Department of Astronomy, University of Cape Town, Private Bag X3, Rondebosch 7701, South Africa\label{uct}
\and
Astronomical Observatory, University of Warsaw, Al. Ujazdowskie 4, 00-478, 
Warszawa, Poland\label{uw}
\and
Institut f{\"u}r Astronomie und Astrophysik, Sand 1, 72076 T{\"u}bingen, Germany\label{iaat}
\and
Remeis Observatory and ECAP, Universit{\"a}t Erlangen-N{\"u}rnberg, Sternwartstra{\ss}e 7, 96049 Bamberg, Germany\label{uerl}
\and
Universit\'e de Strasbourg, CNRS, Observatoire astronomique de Strasbourg, UMR 7550, 67000 Strasbourg, France\label{oas}
}

\date{Received 28 December 2022 / Accepted 23 January 2023}

\abstract
   {Using data from \ero, the soft X-ray instrument aboard {\it Spectrum-Roentgen-Gamma} (\srg), we report the discovery of two new hard transients, \asrc and \csrc, in the Large Magellanic Cloud. We also report the detection of the Be/X-ray binary \bsrc in a bright state.}
   {We initiated follow-up observations to investigate the nature of the new transients and to search for X-ray pulsations coming from \bsrc.}
   {We analysed the X-ray spectra and light curves from our \xmm observations, obtained optical spectra using the South African Large Telescope to look for Balmer emission lines and utilised the archival data from the Optical Gravitational Lensing Experiment (OGLE) for the long-term monitoring of the optical counterparts.}
   {We find X-ray pulsations for \asrc, \bsrc, and \csrc of 40.6\,s, 17.3\,s, and 784\,s, respectively. The \Halpha emission lines with equivalent widths of -10.4\,\AA\ (\asrc) and -43.9\,\AA\ (\csrc) were measured, characteristic for a circumstellar disc around Be stars. The OGLE I- and V-band light curves of all three systems exhibit strong variability. A regular pattern of deep dips in the light curves of \bsrc suggests an orbital period of $\sim$451\,days.}
   {We identify the two new hard \ero transients \asrc and \csrc and the known Be/X-ray binary \bsrc as Be/X-ray binary pulsars.}


\maketitle   


\section{Introduction}
\label{sec:intro}

 The soft X-ray instrument \ero on board the {\it Spektrum-Roentgen-Gamma} (\srg) mission \citep{2021A&A...647A...1P} began scanning the sky in great circles in December 2019. Until December 2021, one full \ero sky survey  was completed every half-year (eRASS1 to eRASS4), whereas eRASS5 was stopped in February 2022.
The survey strategy with scans along great circles (six per day), which intersect at the ecliptic north and south poles leads to a higher number of scans across sources near the poles.
The Large Magellanic Cloud (LMC) is located sufficiently close to the south-ecliptic pole to have the north-eastern side scanned for up to three weeks.

The first scans from eRASS1 imaged the northern part of the LMC, leading to the discovery of \asrc \citep{2020ATel13609....1H} and eRASSU\,J052914.9$-$662446 \citep{2020ATel13610....1M}.
Following its discovery, eRASSU\,J052914.9$-$662446 was observed with \nustar and pulsations in the X-ray flux were discovered \citep{2020ATel13650....1M}. 
The results of a detailed analysis are presented in \citet{2023A&A...669A..30M}.
One year later, \asrc was seen to have brightened and we executed an anticipated target of opportunity (ToO) observation with \xmm, which allowed us to discover X-ray pulsations \citep{2021ATel15133....1H}. The source became brightest at the end of eRASS4/beginning of eRASS5 when it was detected by Mikhail Pavlinsky ART-XC \citep{2021A&A...650A..42P}, the hard X-ray instrument on board \srg. A \nustar observation independently revealed the pulsations of \asrc \citep{2022MNRAS.514.4018S}.

Over the course of eRASS3, when \ero was scanning over \bsrc, the source was found in a bright state, whereas it was not detected during eRASS1 and eRASS2.
In particular, \bsrc is known since it was discovered with the \ein observatory \citep[CAL\,9;][]{1981ApJ...248..925L} and identified as a Be/X-ray binary \citep{1985AJ.....90...43C,1984ApJ...286..196C}. The cited authors suggested the variable star SV*\,HV\,2289 as optical counterpart and found \Hbeta in emission in their blue spectrum which indicates a B0e spectral type.
We again triggered an \xmm ToO observation with the aim to confirm the identification and detect pulsations.

Another new transient X-ray source with a hard spectrum (\csrc) was found during eRASS4 on the western side of the LMC towards the Magellanic Bridge. As a possible optical counterpart, we identified a star with colour and brightness consistent with a B star, suggesting a Be/X-ray binary suitable for an \xmm follow-up observation.
    
In this paper, we describe the results from our X-ray observations
of \asrc, \bsrc, and \csrc, using eROSITA, XMM-Newton,
and Swift. These results are detailed in Sect. \ref{sec:xray}. 
In Sect. \ref{sec:cpart}, we present the identification of the optical counterparts, their long-term monitoring recorded by the Optical Gravitational Lensing Experiment \citep[OGLE;][]{2008AcA....58...69U,2015AcA....65....1U} and optical spectra (Sect. \ref{sec:salt}) obtained
with the South African Large Telescope \citep[SALT;][]{Buckley2006}.
We discuss our final results in Sect. \ref{sec:discussion} and conclusions in Sect. \ref{sec:conclusion}.

\section{X-ray observations and data analysis}
\label{sec:xray}

\subsection{\ero}

To analyse the data, we used the \ero Standard Analysis Software System \citep[\eSASS][]{2022A&A...661A...1B} version {\tt eSASSusers\_211214}.
To produce \ero source products like light curves and spectra, we used the \eSASS task \texttt{srctool}  \citep[see e.g.][]{2021A&A...647A...8M,2022A&A...661A..25H}. We defined circular regions to extract source (radius 50\arcsec, 60\arcsec,\ and 40\arcsec) and background events (from a nearby source-free region; radius 50\arcsec, 90\arcsec,\ and 65\arcsec) for \asrc, \bsrc, and \csrc, respectively. For the spectra and light curves, we selected all valid pixel patterns (\pattern =\,15). For the light curves, we combined the data from all cameras (telescope modules TM 1--7) and 
applied a cut in the fractional exposure of 0.15 (FRACEXP$>$0.15) to exclude data from the edge of the detectors. We created combined spectra using only data from TM 1--4 and 6, the five cameras with an on-chip filter. We note that TM5 and TM7 suffer from a light leak \citep{2021A&A...647A...1P} and no reliable energy calibration is available yet. The \ero source spectra were binned to a minimum of one count per bin to use Cash statistics \citep{1979ApJ...228..939C}. 

The \ero light curves of \asrc, \bsrc, and \csrc are shown in Figs.\,\ref{fig:erolca}, \ref{fig:erolcb}, and \ref{fig:erolcc}.
\asrc is located closest to the south-ecliptic pole and was scanned most often, including the first part of eRASS5 in total 272 times, while \csrc in the west of the LMC was scanned 134 times. The light curves of \asrc and \bsrc show variability by a factor of a few on timescales of weeks and between surveys. \csrc was detected only during eRASS4 and too faint to draw conclusions about variability over days.

The analysis of the \ero spectra (and also \xmm EPIC spectra, see below) was performed using \xspec v12.10.1f \citep{1996ASPC..101...17A}.
We fitted the \ero spectra using a simple absorbed power-law model with two absorption components, 
one accounting for the foreground absorption in the Galaxy with ISM abundances 
following \citet{2000ApJ...542..914W} and atomic cross-sections from \citet{1996ApJ...465..487V}. 
The Galactic column density, \nhGal, was taken from \citet{1990ARA&A..28..215D} and fixed in the fits (4.4, 8.6 and 7.2\hcm{20} for \asrc, \bsrc, and \csrc, respectively).
For the absorption along the line of sight through the LMC and local to the source, we fixed the elemental abundances at 0.49 solar \citep{2002A&A...396...53R,1998AJ....115..605L} and left the column density free in the fit. 
Errors on spectral fit parameters indicate 90\% confidence intervals throughout the paper. 

We started to fit the \ero spectra of \asrc from the five epochs individually. 
The best-fit parameters are assembled in Table\,\ref{tab:eROSpectralFit} (upper part) 
and indicate no significant changes in spectral shape (column density or power-law index) with time.
The overall flux varies by a factor of 6.3 (4.0--13.4). 
Given the constant spectral shape, we next fitted the spectra simultaneously using a common absorption and power-law index (Table\,\ref{tab:eROSpectralFit}, second part).
In this case the flux varies by a factor of 7.4 (5.9--9.5).
The power-law index is typical for HMXBs observed in the 0.2--10 keV band \citep{2008A&A...489..327H} and the X-ray luminosity
varied between 0.5 and 3.6\ergs{36} on the half-year timescale defined by the \ero surveys.

\bsrc was also detected in each of the first four eRASS surveys, but at a fainter flux level. Therefore, we fitted the four \ero spectra simultaneously with common power-law index and absorption. No significant LMC absorption was required in the fit. The best-fit parameters are listed in Table\,\ref{tab:eROSpectralFit}.

\begin{table}
    \centering
    \caption{Analysis of \ero spectra.}
    \label{tab:eROSpectralFit}
    \begin{tabular}{c@{\hspace{3pt}}cc@{\hspace{4pt}}c@{\hspace{4pt}}c}
    \hline\hline\noalign{\smallskip}
    eRASS\tablefootmark{(a)} & Power-law & \nhlmc\tablefootmark{(b)} & F$_{\rm x}$\tablefootmark{(c)} & L$_{\rm x}$\tablefootmark{(d)} \\
          & index     & \oexpo{20}                & \oexpo{-12}                    & \oexpo{36}                     \\
          &           & \uhcm                     & \uergcm                        & \uergs                         \\
    \noalign{\smallskip}\hline\noalign{\smallskip}
    \multicolumn{5}{c}{\asrc} \\
    \noalign{\smallskip}\hline\noalign{\smallskip}
    0--1 & 1.2$\pm0.4$ & 30.0$^{+23.5}_{-17.6}$ & 3.0$^{+1.1}_{-0.8}$ & 1.03$^{+0.28}_{-0.19}$\\ \noalign{\smallskip}
    1--2 & 0.8$\pm0.4$ & 11.5$^{+17.4}_{-11.5}$ & 1.8$^{+0.7}_{-0.6}$ & 0.57$^{+0.20}_{-0.15}$\\ \noalign{\smallskip}
    2--3 & 1.07$^{+0.23}_{-0.20}$ & 17.8$^{+9.5}_{-7.5}$ & 6.1$^{+1.2}_{-1.0}$ & 2.00$^{+0.31}_{-0.26}$\\ \noalign{\smallskip}
    3--4 & 0.98$^{+0.23}_{-0.20}$ & 11.4$^{+10.1}_{-7.6}$ & 4.7$^{+1.0}_{-0.8}$ & 1.52$^{+0.26}_{-0.21}$\\ \noalign{\smallskip}
    4--5 & 0.97$\pm0.13$ & 12.9$^{+5.1}_{-4.5}$ & 11.4$^{+1.3}_{-1.2}$ & 3.7$\pm0.3$\\ \noalign{\smallskip}
    \noalign{\smallskip}\hline\noalign{\smallskip}
    \multicolumn{5}{c}{\asrc simultaneous fit} \\
    \multicolumn{5}{c}{C-statistic = 1216; $\chi^2$ = 1511; d.o.f. = 1510\tablefootmark{(e)}} \\
    \noalign{\smallskip}\hline\noalign{\smallskip}
    0--1 & & & 3.2$^{+0.5}_{-0.4}$ & 1.04$^{+0.14}_{-0.13}$\\ \noalign{\smallskip}
    1--2 &  &  & 1.52$^{+0.24}_{-0.22}$ & 0.49$\pm0.07$\\ \noalign{\smallskip}
    2--3 & 1.00$\pm0.10$ & 14.5$^{+3.8}_{-3.5}$ & 6.4$\pm0.6$ & 2.07$^{+0.18}_{-0.17}$\\ \noalign{\smallskip}
    3--4 & & & 4.7$\pm0.5$ & 1.53$^{+0.15}_{-0.14}$\\  \noalign{\smallskip}
    4--5 & & & 11.2$^{+1.0}_{-0.9}$ & 3.63$^{+0.27}_{-0.25}$\\ \noalign{\smallskip}
    \noalign{\smallskip}\hline\noalign{\smallskip}
    \multicolumn{5}{c}{\bsrc simultaneous fit} \\
    \multicolumn{5}{c}{C-statistic = 352.6; $\chi^2$ = 363.3; d.o.f. = 426\tablefootmark{(e)}} \\
    \noalign{\smallskip}\hline\noalign{\smallskip}
    1 & & & 0.10$^{+0.07}_{-0.06}$ & 0.031$^{+0.024}_{-0.019}$\\ \noalign{\smallskip}
    2 & \multirow[c]{2}{*}{1.10$^{+0.24}_{-0.14}$} & \multirow[c]{2}{*}{<7} & 0.22$^{+0.11}_{-0.09}$ & 0.070$^{+0.035}_{-0.028}$\\ \noalign{\smallskip}
    3 & & & 2.05$^{+0.37}_{-0.41}$ & 0.66$^{+0.13}_{-0.11}$\\ \noalign{\smallskip}
    4 & & & 2.60$^{+0.48}_{-0.50}$ & 0.83$^{+0.07}_{-0.13}$\\ \noalign{\smallskip}
    \noalign{\smallskip}\hline\noalign{\smallskip}
    \multicolumn{5}{c}{\csrc eRASS4} \\
    \multicolumn{5}{c}{C-statistic = 76.2; $\chi^2$ = 82.9; d.o.f. = 89\tablefootmark{(e)}} \\
    \noalign{\smallskip}\hline\noalign{\smallskip}
    4 & 0.7$^{+1.0}_{-0.6}$ & <102 & 1.8$^{+1.2}_{-0.9}$ & 0.59$^{+0.34}_{-0.20}$\\
    \noalign{\smallskip}\hline
    \end{tabular}
    \tablefoot{
    \tablefoottext{a}{For the definition of the observing epochs see the Appendix.}
    \tablefoottext{b}{The total absorption column consists of \nhGal, which is fixed to the Galactic foreground with solar abundances, 
                      and \nhlmc, which accounts for the additional absorption component within the LMC.}
    \tablefoottext{c}{Observed flux in the energy range of 0.2--8\,keV.}
    \tablefoottext{d}{Corresponding unabsorbed luminosity assuming a source distance of 50\,kpc.}
    \tablefoottext{e}{Best-fit Cash statistics with corresponding $\chi^2$ value and degrees of freedom (d.o.f.).}}
\end{table}

\begin{figure}[ht!]
\centering
 \resizebox{0.95\hsize}{!}{\includegraphics{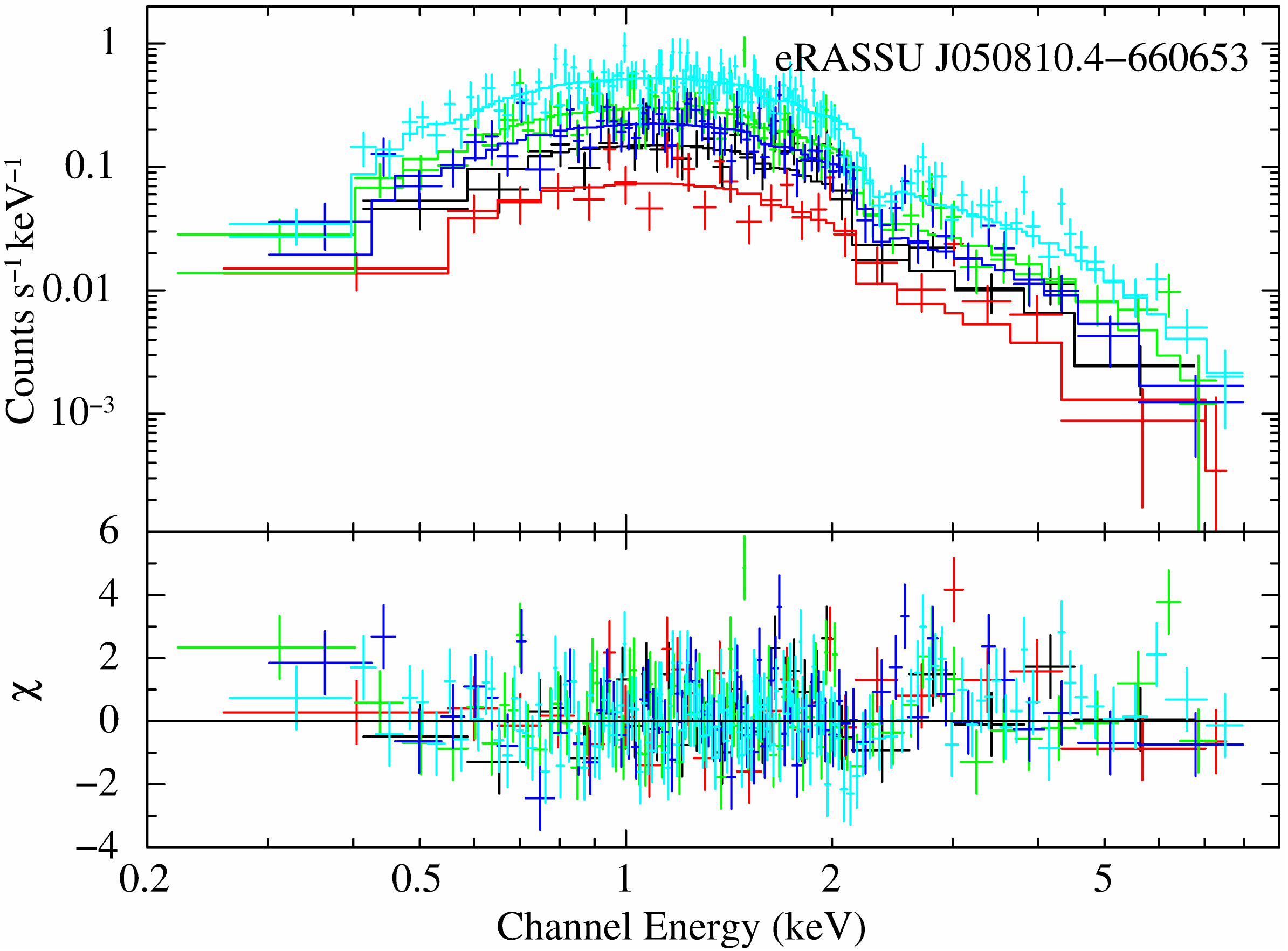}}
 \resizebox{0.95\hsize}{!}{\includegraphics{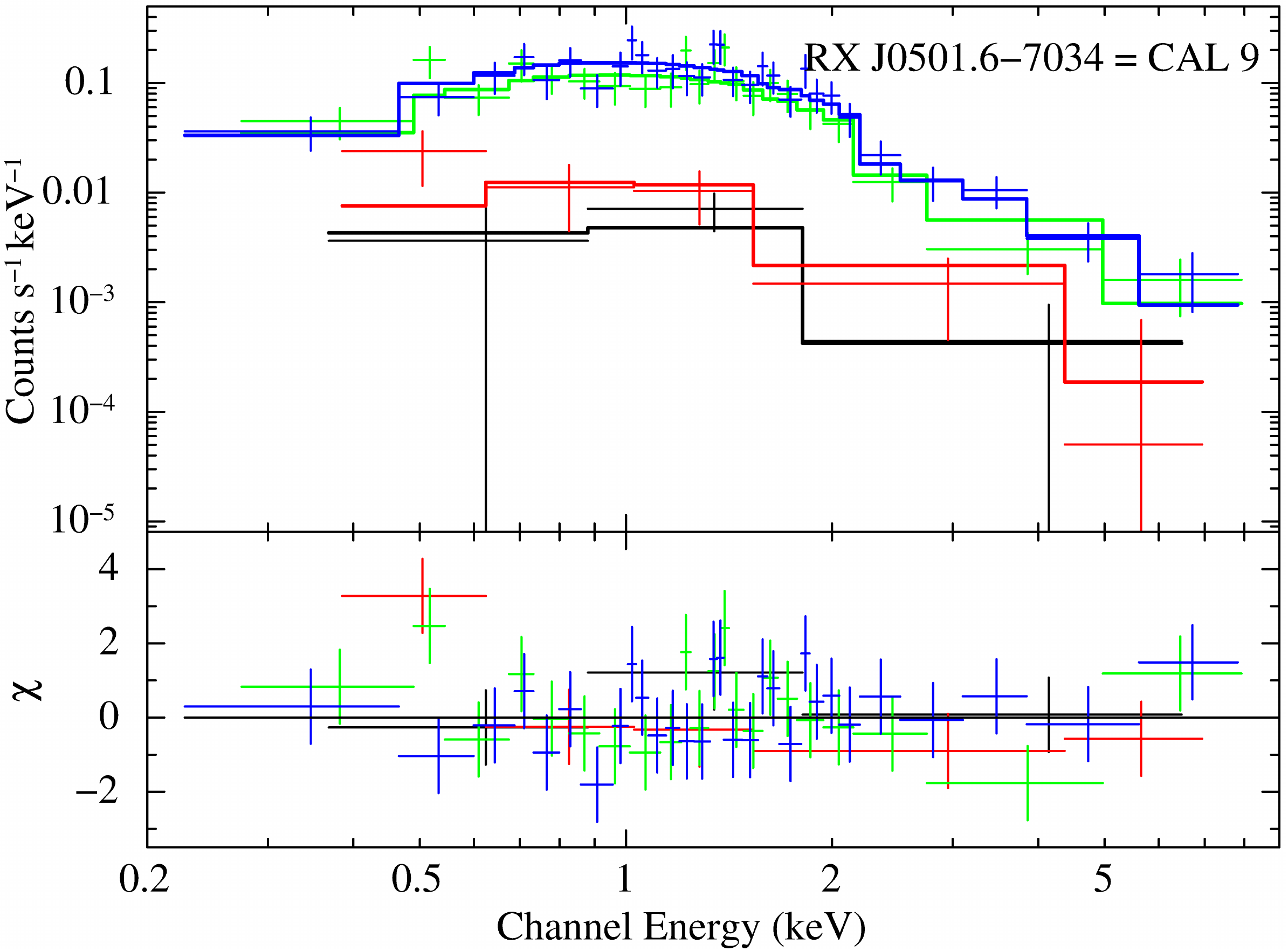}}
 \resizebox{0.95\hsize}{!}{\includegraphics{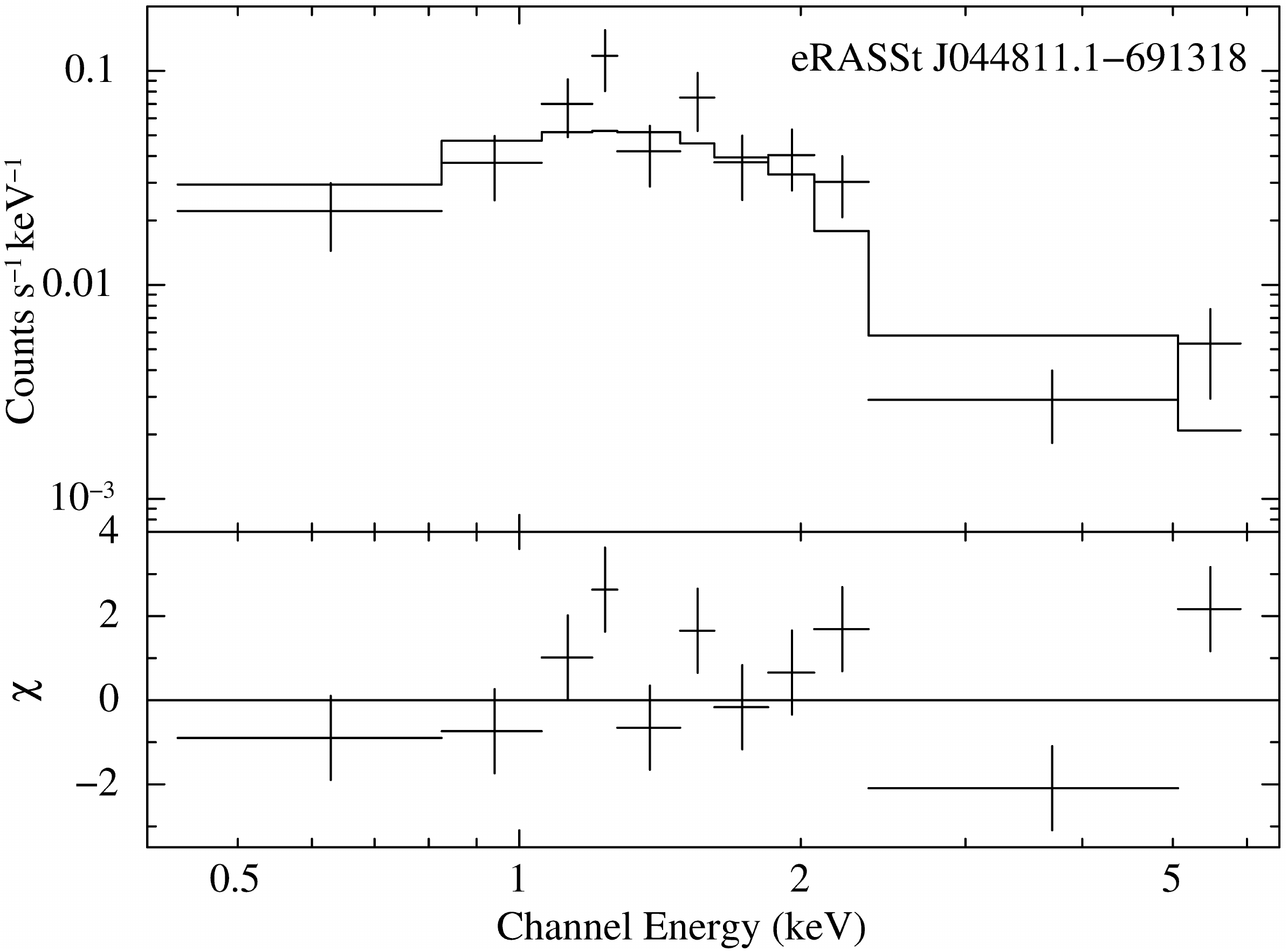}}
  \caption{\ero spectra of \asrc ({\it top}), \bsrc ({\it middle}), and \csrc (bottom) together with their best-fit models (as histograms). The respectively lower panels show the residuals from the best-fit model divided by the 1$\sigma$ errors. For \asrc, the simultaneous spectral fit to the spectra of the different epochs defined in Fig.\,\ref{fig:erolca} is shown (epoch 1 marked in black, epoch 2 in red, epoch 3 in green, epoch 4 in blue, and epoch 5 in turquoise. For \bsrc the simultaneous fit to the spectra of eRASS1 (black), eRASS2 (red), eRASS3 (green), and eRASS4 (blue) is presented. Finally, the eRASS4 spectrum of \csrc is shown at the bottom.}
  \label{fig:erospec}
\end{figure}

\csrc was only detected during eRASS4 and we used the corresponding spectrum for our spectral analysis (Table\,\ref{tab:eROSpectralFit}). Although it is not very well constrained, the derived power-law index indicates a harder X-ray spectrum compared to the other two sources. Again, no significant LMC absorption was required, but column densities up to \ohcm{22} cannot be ruled out. The luminosity was at a similar level as seen from \bsrc and at the lower luminosity range covered by \asrc.
The \ero spectra with their best-fit models are plotted in Fig.\,\ref{fig:erospec}.



\subsection{\xmm}

We triggered \xmm follow-up observations from our anticipated ToO programs for the three targets (see Table\,\ref{tab:xmmobs}).
We used the \xmm Science Analysis Software (SAS)\,19.1.0\footnote{\url{https://www.cosmos.esa.int/web/xmm-newton/sas}}
package to process the data from the European Photon Imaging Camera (EPIC), which is equipped with pn- \citep{2001A&A...365L..18S} and 
MOS-type \citep{2001A&A...365L..27T} charge-coupled device (CCD) detectors.
From the processed EPIC event files, we created such products as images, spectra, and light curves using the SAS task \texttt{evselect}. 

We performed a maximum likelihood source detection simultaneously on the 15 images from the three EPIC instruments in five energy bands (0.2--0.5\,keV, 0.5--1\,keV, 1--2\,keV, 2--4.5\,keV, and 4.5--12\,keV) as described in \citet{2013A&A...558A...3S}. 
The resulting X-ray positions are listed in Table\,\ref{tab:xmmobs}.

We extracted energy spectra from circular regions around source (radius 40\arcsec, 45\arcsec,\ and 30\arcsec) and nearby background (radius 60\arcsec, 60\arcsec,\ and 50\arcsec) for \asrc, \bsrc, and \csrc, respectively. 
Events with \pattern 1--4 for EPIC-pn \citep{2001A&A...365L..18S} and \pattern 1--12 for EPIC-MOS \citep{2001A&A...365L..27T} were selected,
applying the conservative event filtering with \flag 0 for EPIC-pn and EPIC-MOS. 
The EPIC source spectra were binned to a minimum of 20 counts per bin to use Gaussian statistics and the response files were computed with the SAS tasks \arfgen and \rmfgen.

\begin{figure}
\centering
 \resizebox{0.95\hsize}{!}{\includegraphics{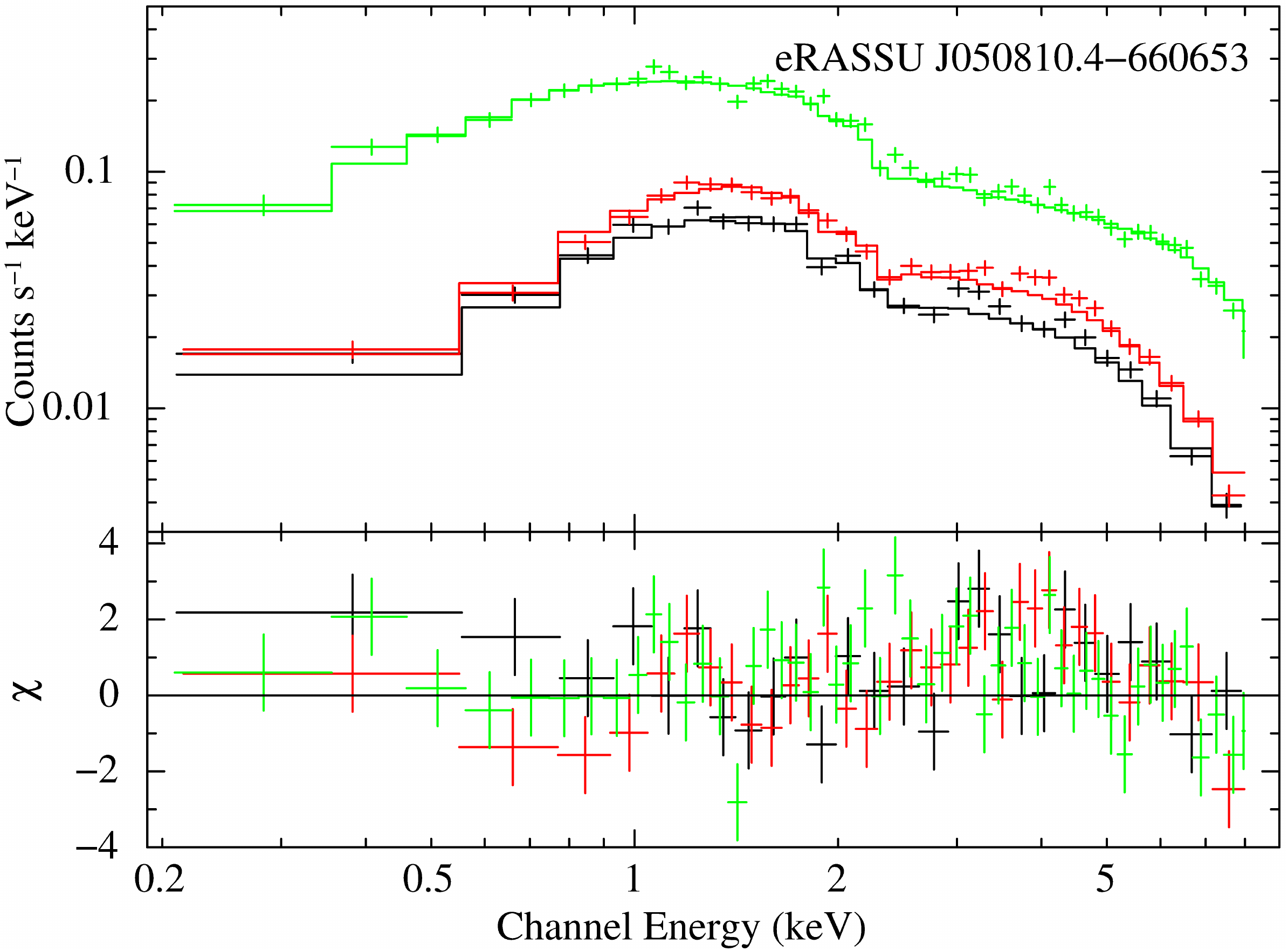}}
 \resizebox{0.95\hsize}{!}{\includegraphics{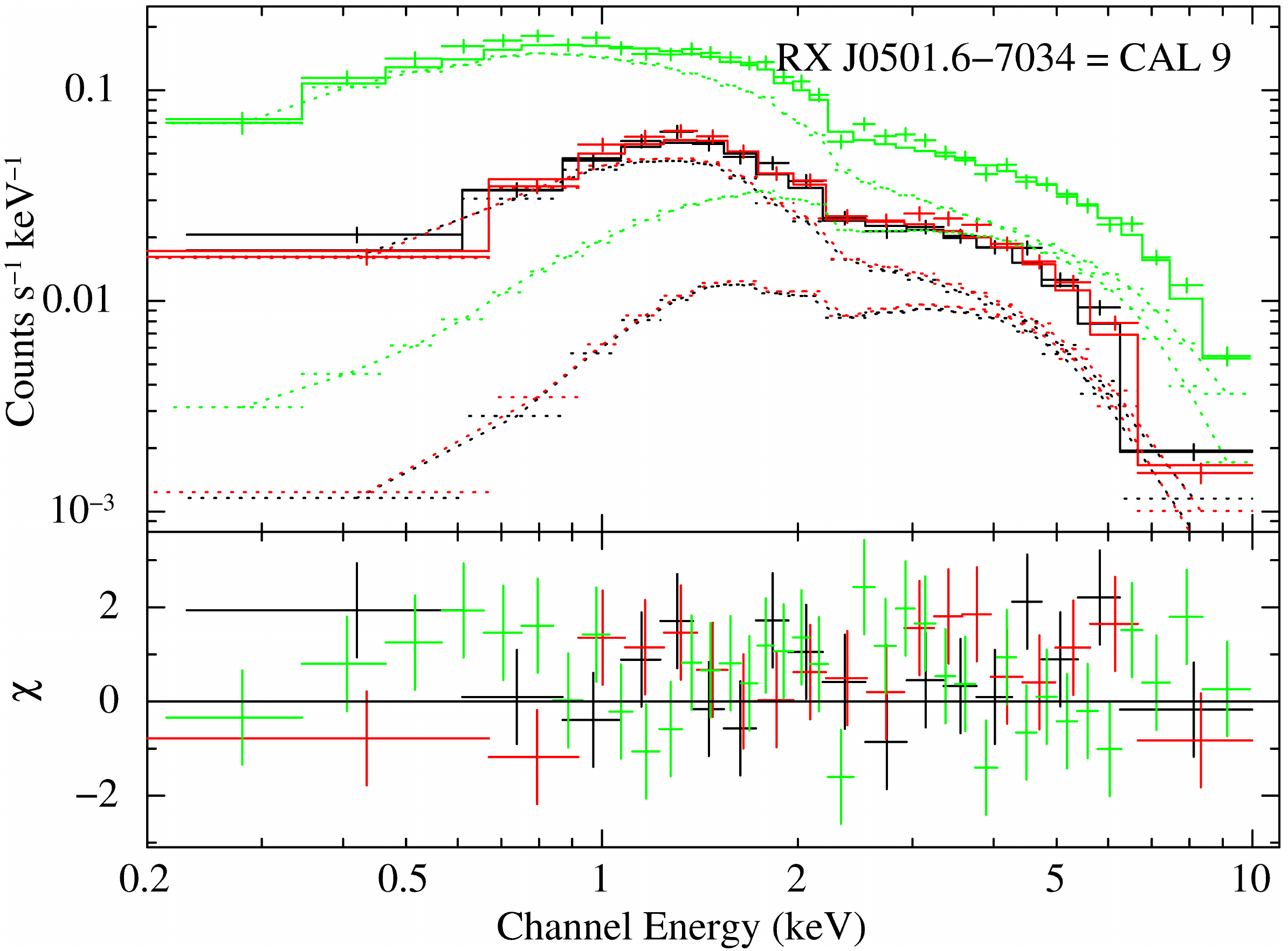}}
 \resizebox{0.95\hsize}{!}{\includegraphics{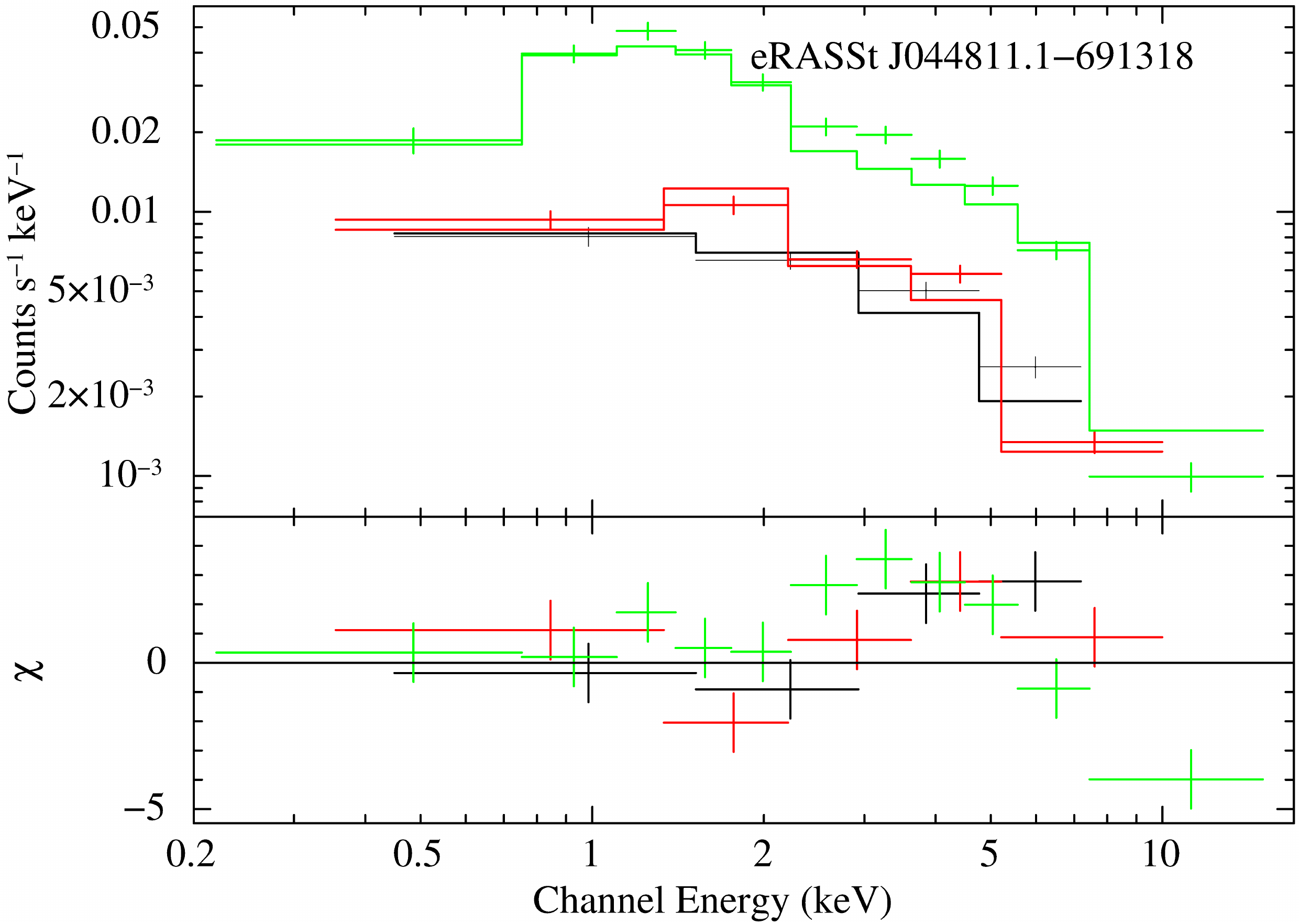}}
  \caption{
    EPIC spectra of the three new Be/X-ray binary pulsars (MOS1: black, MOS2: red, pn: green). 
    The best-fit models are plotted as histogram and the bottom panels show the residuals. For \asrc and \csrc an absorbed power law yielded an acceptable fit; however, for \bsrc, an additional black-body component was included (individual components are indicated by dotted lines). See Table\,\ref{tab:EPresults}, for the model parameters.
  }
  \label{fig:EPspectra}
\end{figure}

In modelling the EPIC spectra, we followed the same approach as for the \ero spectra with a power law with two absorption column densities as basic model. This model adequately fits the spectra from \asrc. However, the residuals from the fit to the spectra of \bsrc reveal a systematic pattern that indicates a second emission component. We added a black-body component which was also used for other BeXRBs \citep[e.g.][]{2022A&A...662A..22H}. Also, the fit to the spectra of \csrc suggests that the pure power-law model is too simple, but the low statistical quality of the spectrum prevents a more detailed spectral study. The derived best-fit parameters for the three sources are summarised in Table\,\ref{tab:EPresults} and the spectra with best-fit model are shown in Fig.\,\ref{fig:EPspectra}.

\begin{table*}
\centering
\caption[]{\xmm observations of \asrc, \bsrc, and \csrc.}
\label{tab:xmmobs}
\begin{tabular}{l@{\hspace{4pt}}c@{\hspace{4pt}}lcr@{\hspace{5pt}}r@{\hspace{5pt}}r}
\hline\hline\noalign{\smallskip}
\multicolumn{1}{c}{Source} &
\multicolumn{1}{c}{Obs.} &
\multicolumn{1}{c}{Observation} &
\multicolumn{1}{c}{Exposure time} &
\multicolumn{1}{c}{R.A.} &
\multicolumn{1}{c}{Dec.} &
\multicolumn{1}{c}{Err} \\
\multicolumn{1}{c}{name} &
\multicolumn{1}{c}{ID} &
\multicolumn{1}{c}{time} &
\multicolumn{1}{c}{pn, MOS1, MOS2} &
\multicolumn{2}{c}{(J2000)} &
\multicolumn{1}{c}{1$\sigma$} \\
\multicolumn{1}{c}{} &
\multicolumn{1}{c}{} &
\multicolumn{1}{c}{} &
\multicolumn{1}{c}{(s)} &
\multicolumn{1}{c}{(h m s)} &
\multicolumn{1}{c}{(\degr\ \arcmin\ \arcsec)} &
\multicolumn{1}{c}{(\arcsec)} \\
\noalign{\smallskip}\hline\noalign{\smallskip}
\asrc & 0860800301 & 2020-12-17 09:54 -- 19:24  & 19500, 31779, 32332 & 05 08 09.99 & -66 06 51.6 & 0.51 \\
\noalign{\smallskip}
\bsrc (CAL\,9) & 0883950101 & 2021-05-18 00:30 -- 08:57  & 20300, 24743, 24617 & 05 01 23.66 & -70 33 34.0 & 0.51 \\
\noalign{\smallskip}
\csrc & 0883950201 & 2021-11-29 01:59 -- 08:22 & 16383, 21075, 21424 & 04 48 10.67 & -69 13 18.0 & 0.52 \\
\noalign{\smallskip}\hline
\end{tabular}
\tablefoot{
The net exposure times after background-flare screening are listed for pn, MOS1 and MOS2, respectively. 
The positional errors include the statistical (1$\sigma$) and a systematic error of 0.5\arcsec\ \citep[see][]{2013A&A...558A...3S}, added quadratically.
}
\end{table*}

\begin{table*}
\centering
\caption[]{EPIC X-ray timing and spectral analysis results.}
\label{tab:EPresults}
\begin{tabular}{lccccccc}
\hline\hline\noalign{\smallskip}
\multicolumn{1}{c}{Source} &
\multicolumn{1}{c}{Period\tablefootmark{a}} &
\multicolumn{1}{c}{\nhlmc} &
\multicolumn{1}{c}{Photon} &
\multicolumn{1}{c}{kT} &
\multicolumn{1}{c}{Flux\tablefootmark{b}} &
\multicolumn{1}{c}{L$_{\rm x}$\tablefootmark{b,c}} &
\multicolumn{1}{c}{$\chi^2_r$/dof}\\
\multicolumn{1}{c}{(short)} &
\multicolumn{1}{c}{(s)} &
\multicolumn{1}{c}{(\ohcm{20})} &
\multicolumn{1}{c}{index} &
\multicolumn{1}{c}{(keV)} &
\multicolumn{1}{c}{(erg cm$^{-2}$ s$^{-1}$)} &
\multicolumn{1}{c}{(erg s$^{-1}$)} &
\multicolumn{1}{c}{} \\
\noalign{\smallskip}\hline\noalign{\smallskip}
\sasrc & 40.602544 $\pm$ 7.2\expo{-5} & 8.0 $\pm$ 1.5 & 0.76 $\pm$ 0.02                  & -- & 5.62\expo{-12} & 1.74\expo{36} & 0.96/1107\\
\noalign{\smallskip}
\sbsrc\tablefootmark{d} &  17.3321  $\pm$ 3.8\expo{-3} & $<$2.5 & 1.20$^{+0.12}_{-0.07}$ & 1.61$^{+0.17}_{-0.13}$ & 2.88\expo{-12} & 9.01\expo{35} & 1.00/729\\
\noalign{\smallskip}
\scsrc &  783.75 $\pm$ 0.55           & 13.4 $\pm$ 6.2 & 0.80 $\pm$ 0.06                 & -- & 1.02\expo{-12} & 3.19\expo{35} & 1.40/186\\
\noalign{\smallskip}\hline
\end{tabular}
\tablefoot{
\tablefoottext{a}{Most probable period with 1$\sigma$ error.}
\tablefoottext{b}{X-ray flux and luminosity in the 0.2--10\,keV band.}
\tablefoottext{c}{Absorption-corrected luminosity assuming a distance of 50\,kpc.}
\tablefoottext{d}{
To model the spectra of \bsrc an additional black-body component was included.
It contributes 36.7\% to the total flux and 35.3\% to the luminosity.
The $\chi^2_r$ for the fit without the black-body component was 1.10 for 731 degrees of freedom (dof).
The F-test probability of 3.2\expo{-16} justifies the addition of this component.
}
}
\end{table*}
\begin{table*}
\centering
\caption[]{Optical counterparts of \asrc, \bsrc, and \csrc.}
\label{tab:optical}
\begingroup
\setlength{\tabcolsep}{4pt} 
\begin{tabular}{lccccccccc}
\hline\hline\noalign{\smallskip}
\multicolumn{1}{c}{Source} &
\multicolumn{1}{c}{V\tablefootmark{a}} &
\multicolumn{1}{c}{Q\tablefootmark{a,b}} &
\multicolumn{1}{c}{2MASS} &
\multicolumn{1}{c}{J} &
\multicolumn{1}{c}{H} &
\multicolumn{1}{c}{K$_{\rm s}$} &
\multicolumn{1}{c}{R.A.} &
\multicolumn{1}{c}{Dec.} &
\multicolumn{1}{c}{D\tablefootmark{d}} \\
\multicolumn{1}{c}{(short)} &
\multicolumn{6}{c}{} &
\multicolumn{2}{c}{(J2000)\tablefootmark{c}} &
\multicolumn{1}{c}{} \\
\multicolumn{1}{c}{} &
\multicolumn{1}{c}{(mag)} &
\multicolumn{1}{c}{(mag)} &
\multicolumn{1}{c}{} &
\multicolumn{1}{c}{(mag)} &
\multicolumn{1}{c}{(mag)} &
\multicolumn{1}{c}{(mag)} &
\multicolumn{1}{c}{(h m s)} &
\multicolumn{1}{c}{(\degr\ \arcmin\ \arcsec)} &
\multicolumn{1}{c}{(\arcsec)} \\
\noalign{\smallskip}\hline\noalign{\smallskip}
\sasrc & 14.15 & -0.89 & 05080993$-$6606523 & 13.98 & 13.93 & 13.70 & 05 08 09.94 & -66 06 52.1 & 0.61 \\
\noalign{\smallskip}
\sbsrc & 14.36 & -0.84 & 05012419$-$7033346 & 13.98 & 13.17 & 13.06 & 05 01 23.84 & -70 33 33.9 & 0.95 \\
\noalign{\smallskip}
\scsrc & 15.77 & -0.76 & --                 & --    & --    & --    & 04 48 10.65 & -69 13 17.0 & 0.99 \\
\noalign{\smallskip}\hline
\end{tabular}
\endgroup
\tablefoot{
\tablefoottext{a}{Magnitudes and colours are taken from \citet{2004AJ....128.1606Z} or in the case of \csrc from \citet{2002ApJS..141...81M}.}
\tablefoottext{b}{Reddening-free parameter, defined as Q = U-B - 0.72(B-V). For the distribution of the Q parameter of BeXRBs in the SMC;  see \citet{2016A&A...586A..81H}.}
\tablefoottext{c}{Position of the optical counterpart from Gaia EDR3 \citep[see][]{2021A&A...649A...1G,2016A&A...595A...1G}.}
\tablefoottext{d}{Angular distance between \xmm and Gaia position.}
}
\end{table*}

\begin{figure}
  \centering
    \resizebox{0.96\hsize}{!}{\includegraphics{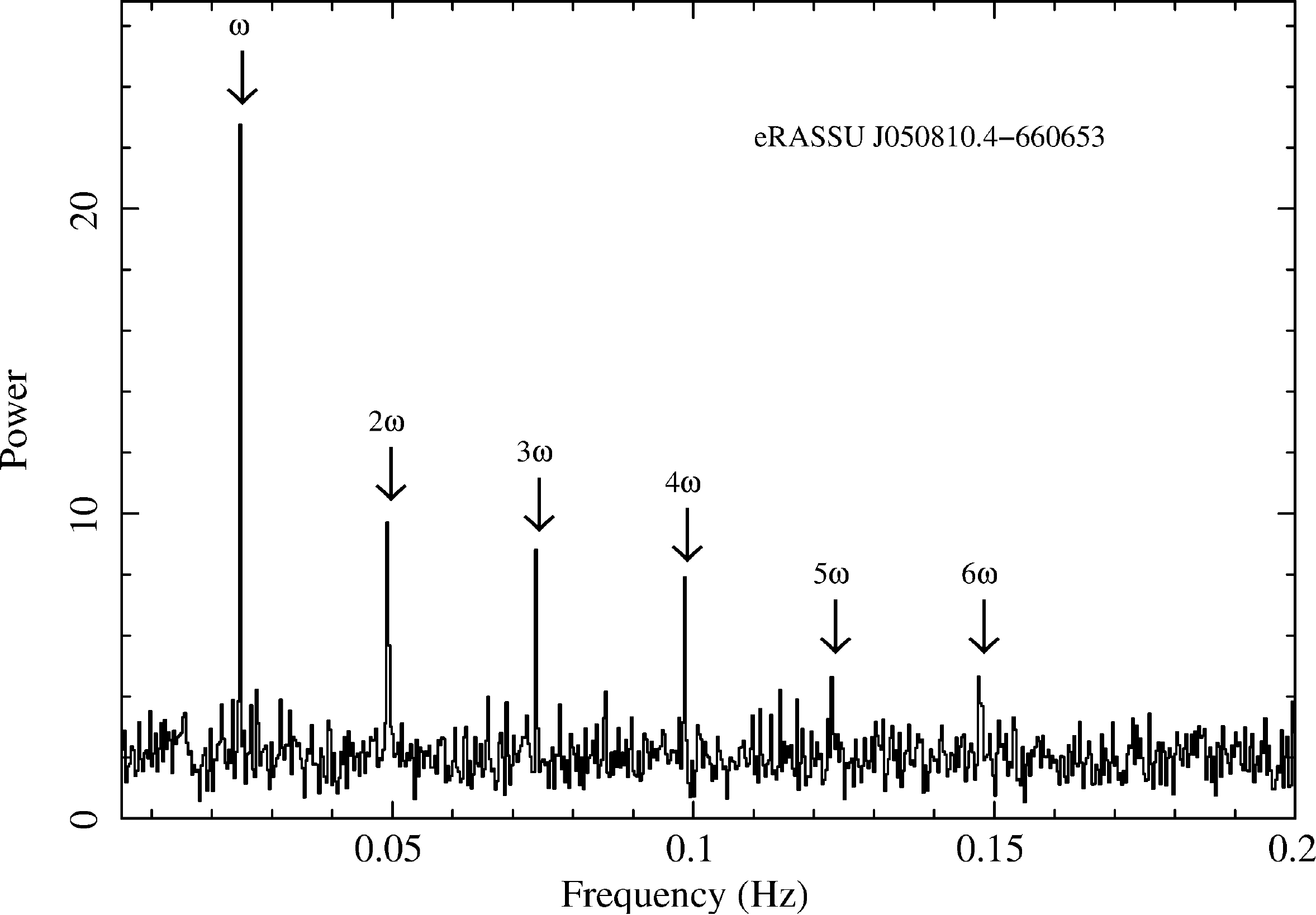}}
    \resizebox{0.96\hsize}{!}{\includegraphics{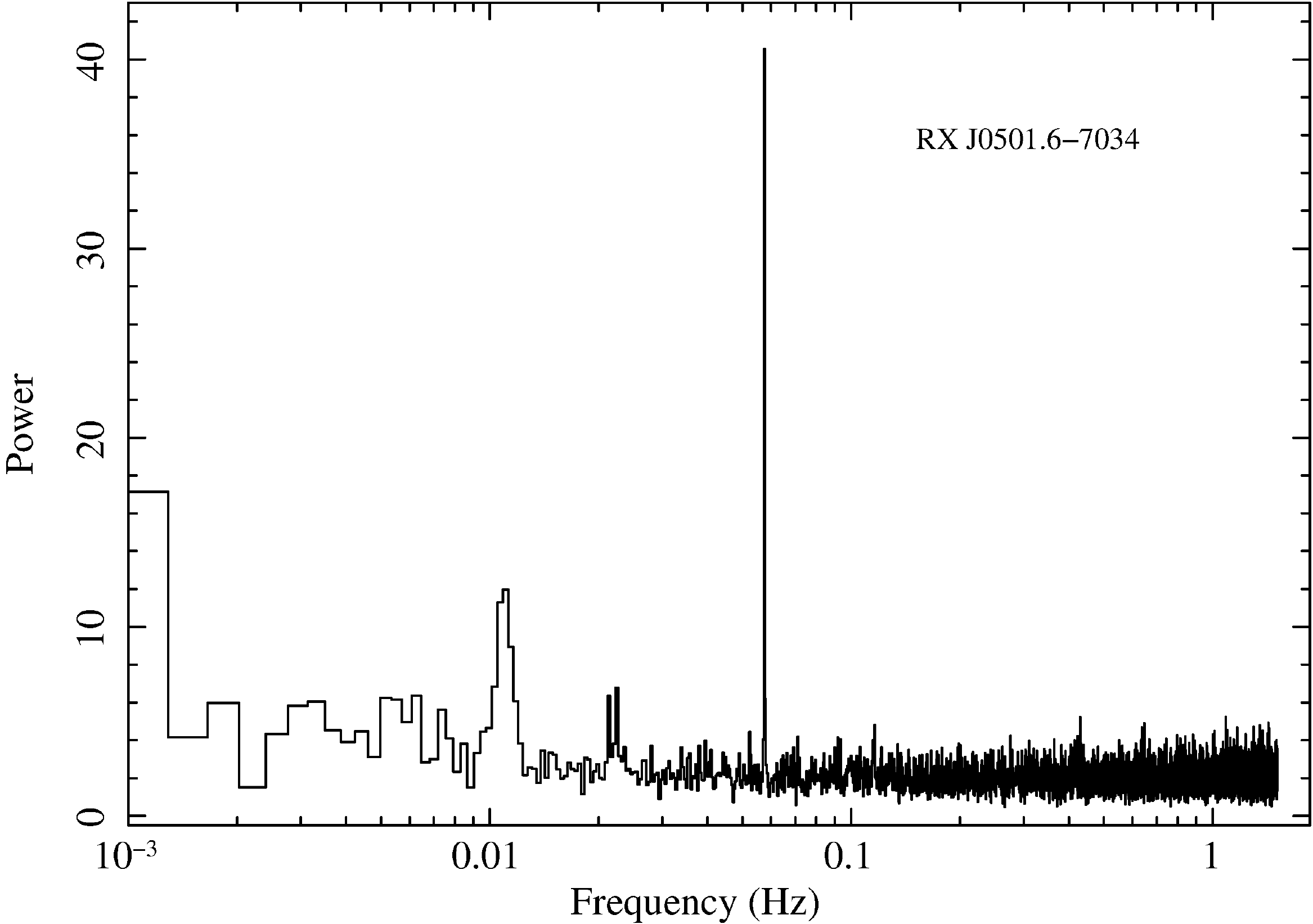}}
    \resizebox{0.96\hsize}{!}{\includegraphics{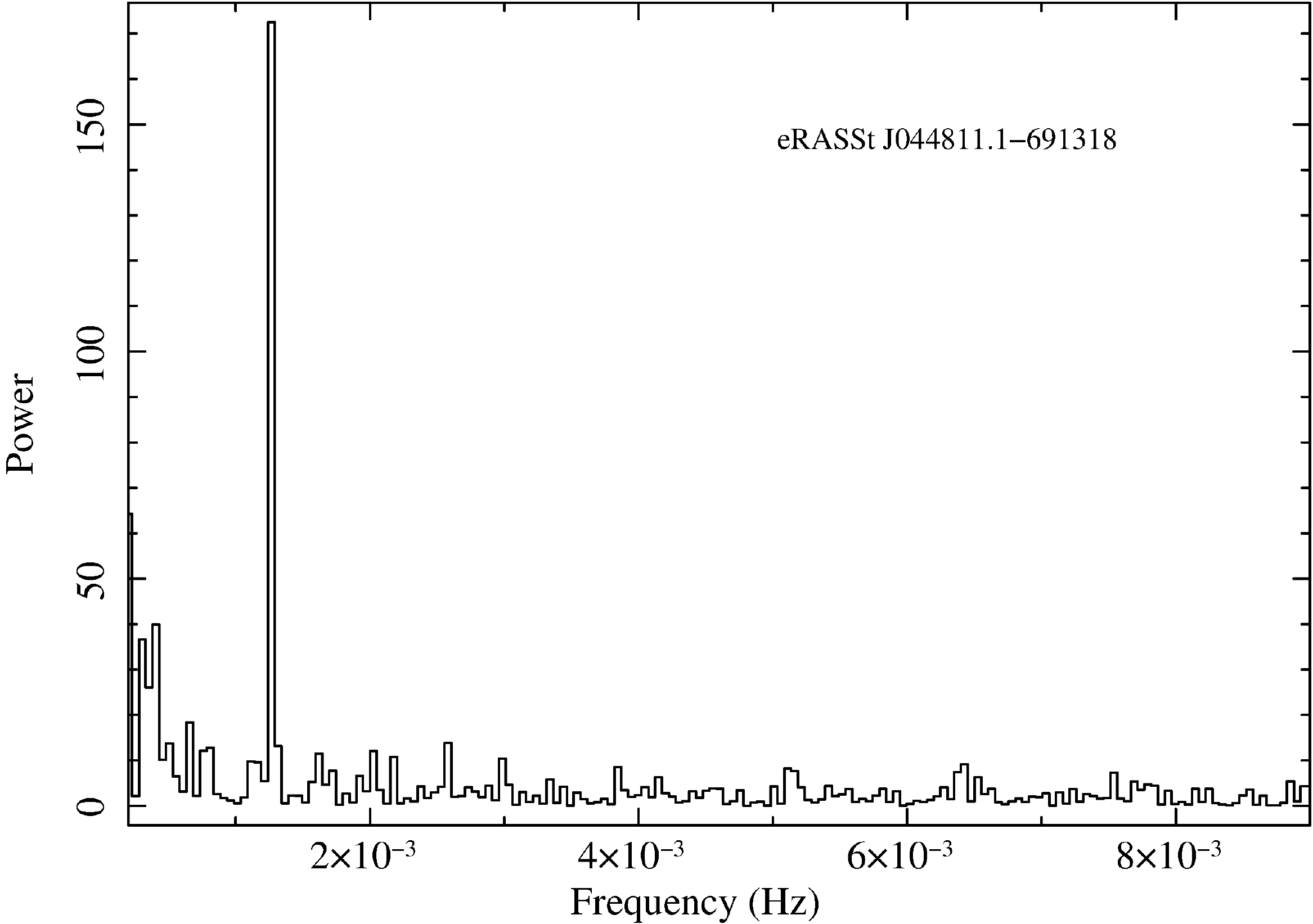}}
  \caption{
    Power spectra obtained from the EPIC-pn data (0.2--8.0\,keV).
    {\it Top:}
    Pulsations from \asrc are clearly detected with a fundamental frequency of 0.0246\,Hz corresponding to a period of 40.60\,s 
    together with signals at five harmonic frequencies.
    {\it Middle:} \bsrc shows pulsations with a frequency of about 57.7\,mHz (17.3\,s).
    {\it Bottom:}  Longest period is detected from \csrc at $\sim$784\,s (frequency of 1.28\,mHz).
  }
  \label{fig:PNpower}
\end{figure}

Correcting the event arrival times to the solar system barycentre and relaxing the selection to valid pixel patterns (\pattern 1--12), we created EPIC-pn light curves in the 0.2--8.0\,keV energy band from the same extraction regions as used for the spectra.
For our search for X-ray pulsations in the light curves, we first created power spectra which revealed pulsations for all our three targets. 
The power spectra obtained from the 0.2--8.0\,keV EPIC-pn light curves are presented in Fig.\,\ref{fig:PNpower}.
The power spectrum of \asrc with five harmonics of the fundamental frequency indicates a complex pulse profile. The power spectra of the other two pulsars, \bsrc and \csrc, only reveal their fundamental frequencies.

To determine the precise period and its error we applied in a second step the Bayesian approach described by \citet{1996ApJ...473.1059G} in a restricted frequency range around the fundamental frequency, which was determined from the power spectrum. 
This method was used already for other HMXB pulsars in the Magellanic Clouds \citep[e.g.][]{2022A&A...662A..22H,2017MNRAS.470.4354V,2013A&A...558A..74V,2008A&A...489..327H}. 

\begin{figure}
\centering
 \resizebox{0.9\hsize}{!}{\includegraphics{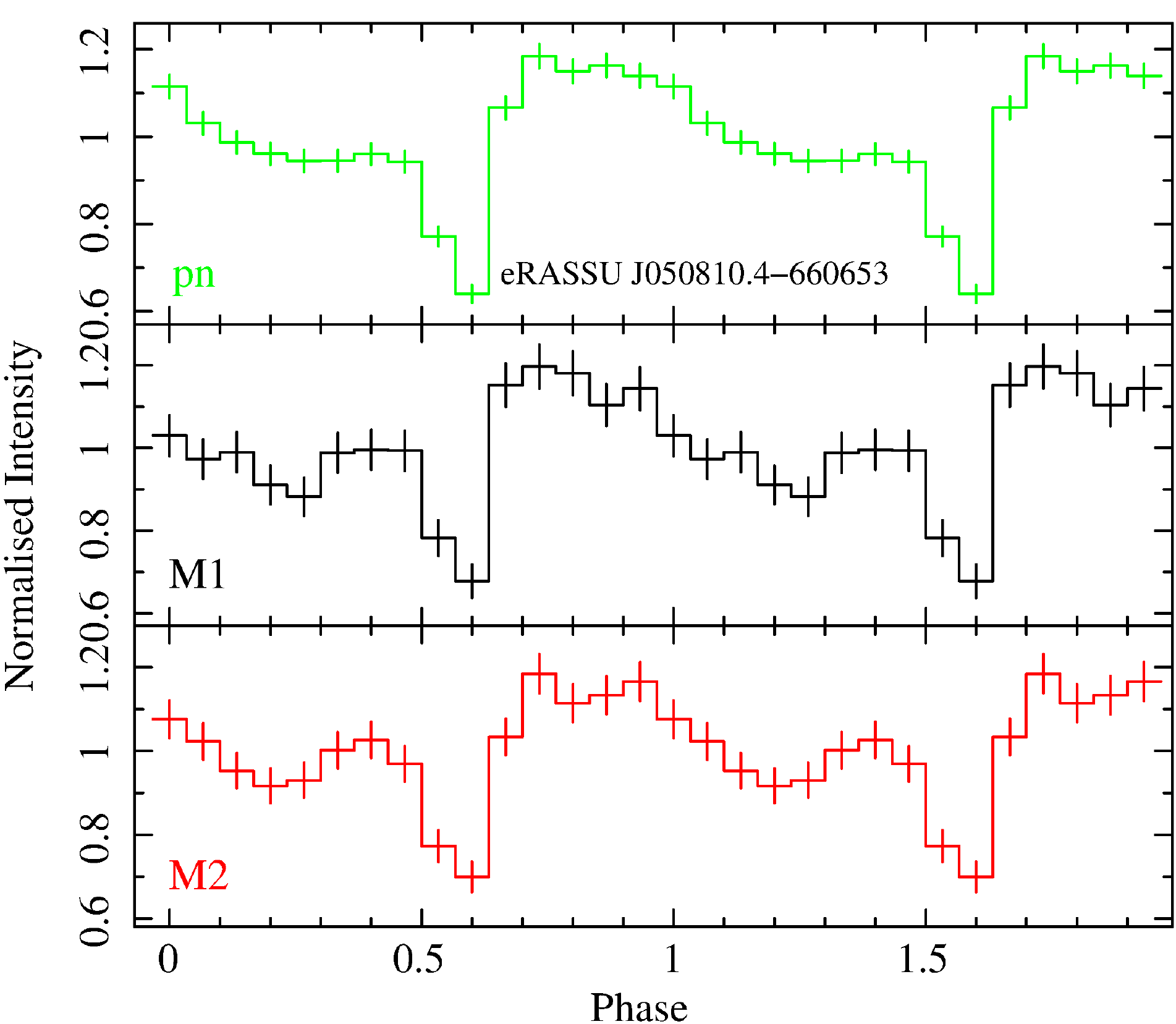}}
 \resizebox{0.9\hsize}{!}{\includegraphics{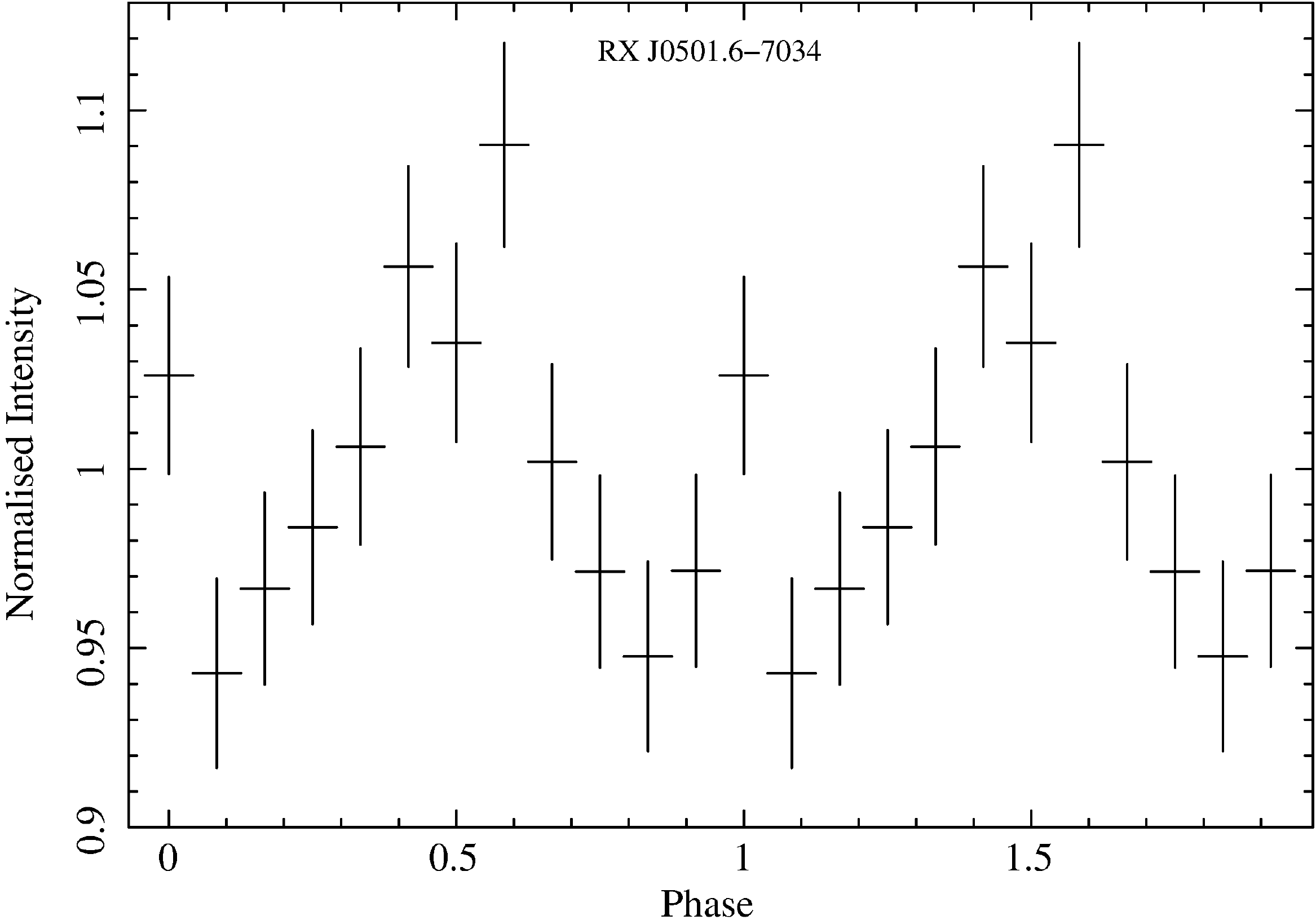}}
 \resizebox{0.9\hsize}{!}{\includegraphics{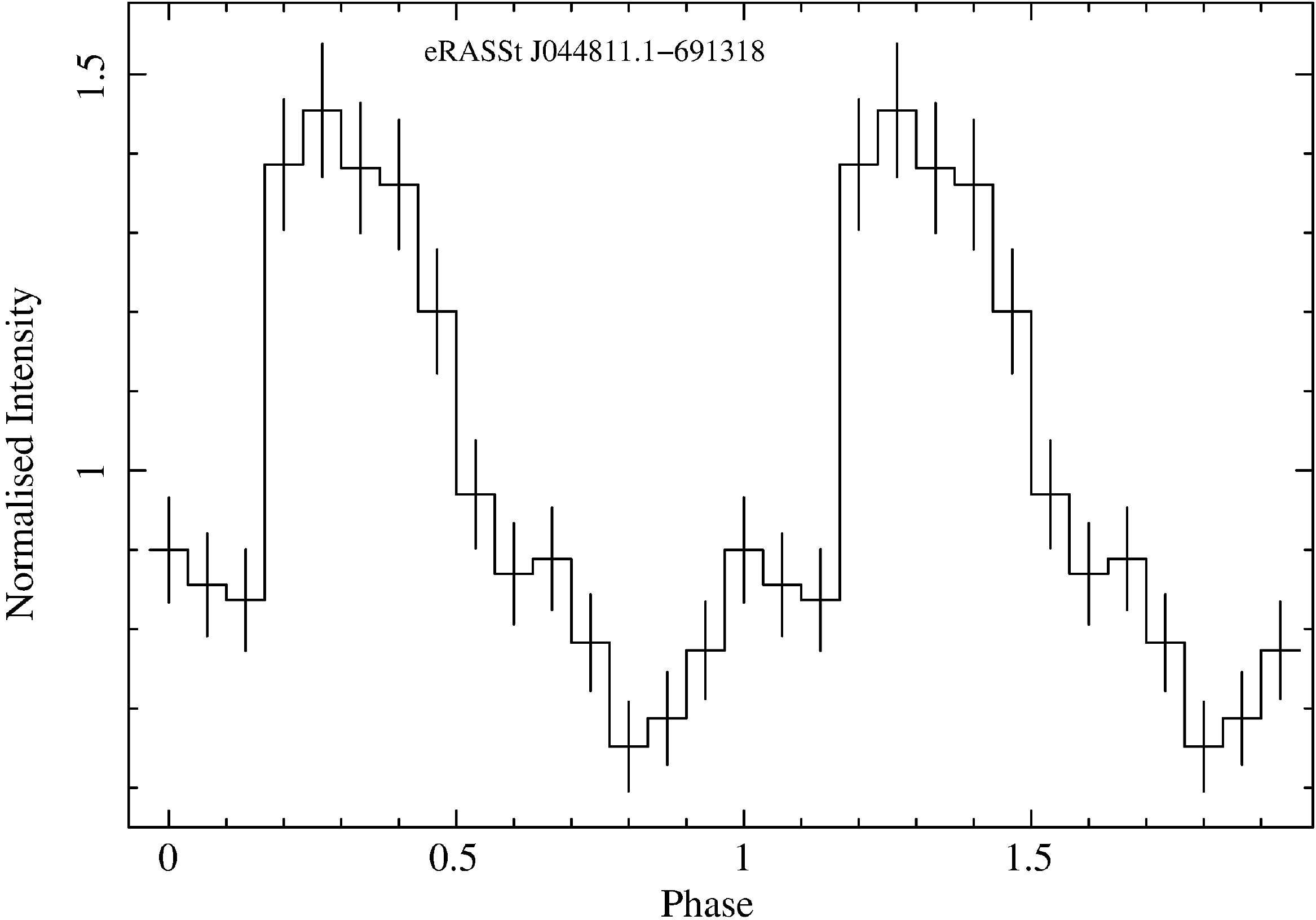}}
  \caption{
    Pulse profiles of our three targets obtained from the EPIC instruments in the 0.2--8.0\,keV band.
    {\it Top:} For \asrc, the profiles from all three EPIC instruments (MOS1: black, MOS2: red, pn: green) are compared.
    The pulse profiles from \bsrc ({\it middle}) and \csrc ({\it bottom}) were derived from EPIC-pn data.
  }
  \label{fig:EPulse}
\end{figure}

The derived periods of the three new pulsars with their 1$\sigma$ errors are listed in Table\,\ref{tab:EPresults}. We folded the light curves with the most probable period and show the pulse profiles in Fig.\,\ref{fig:EPulse}. The pulse profile of \asrc is highly structured with a deep minimum. This dip is barely resolved in time and lasts \lsim\,0.15 in phase. The statistical quality of the pulse profile from \bsrc is low. It is characterised by a main peak of triangular shape and an indication for a second smaller peak. Similarly, \csrc shows a main peak with fast rise preceded by a smaller one.

\section{Optical counterparts}
\label{sec:cpart}

 We identified stars with V-magnitudes between 14.1 and 15.8 as optical counterparts for the three HMXB candidates, based on the values given in the Magellanic Cloud Photometric Survey (MCPS) of \citet{2004AJ....128.1606Z}. 
Two of them also have entries in the Two Micron All Sky Survey \citep[2MASS;][]{2006AJ....131.1163S} catalogue. 
In all three cases, the counterpart is found within the 2-$\sigma$ error circle of the X-ray position (Tables\,\ref{tab:xmmobs} and \ref{tab:optical}). Furthermore, no other star brighter than V=17\,mag and colours compatible with an early-type star is found within 10\arcsec.
The brightness and colours (Table\,\ref{tab:optical}), together with the X-ray spectral and temporal properties strongly suggest a Be/X-ray binary nature for all three objects.

\subsection{SALT spectroscopy}
\label{sec:salt}

We obtained optical spectra from \asrc and \csrc using the Robert Stobie Spectrograph \citep[RSS;][]{2003SPIE.4841.1463B} on SALT.
To investigate the \Halpha line, the PG0900 VPH grating was used, which covered the spectral region 3920--7000\,\AA\ at a resolution of 6.2\,\AA. 
The observations were conducted under the SALT transient follow-up program and are detailed in Table\,\ref{tab:saltobs}.

\begin{figure}
\centering
 \resizebox{\hsize}{!}{\includegraphics{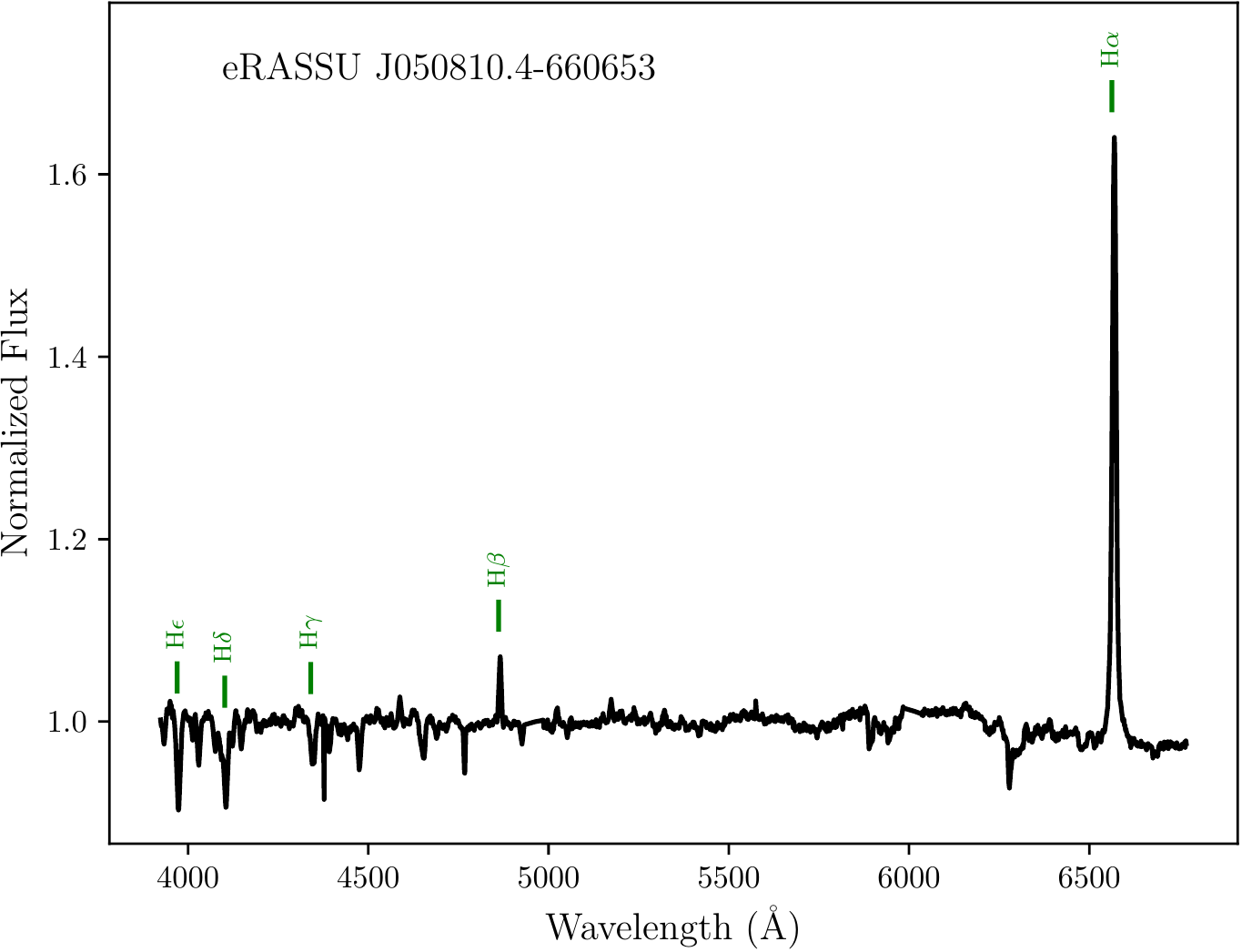}}
 \resizebox{\hsize}{!}{\includegraphics{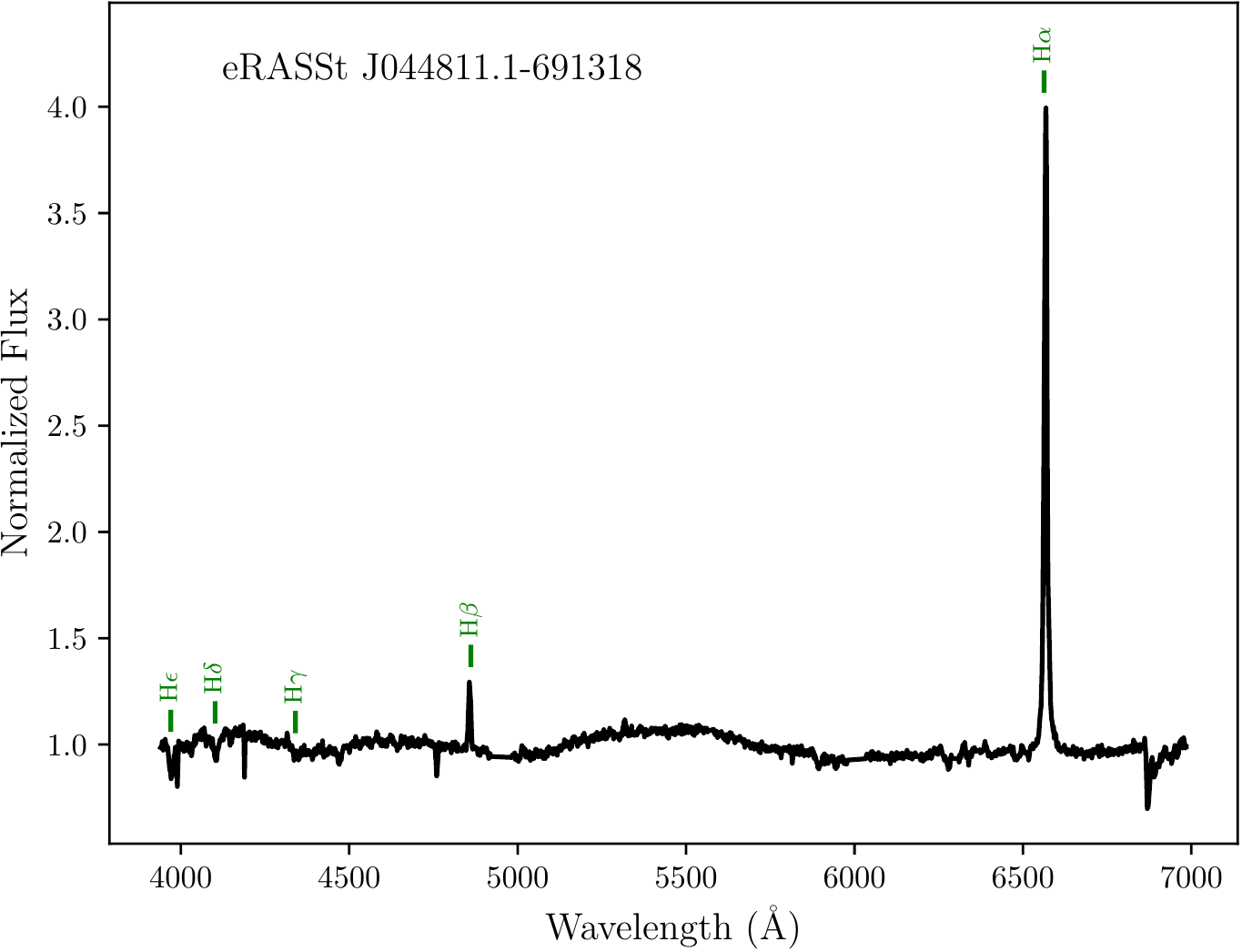}}
  \caption{
    Optical spectra of \asrc (top) and \csrc (bottom) taken with the RSS on SALT. The rest frame wavelengths of the Balmer lines are marked.
  }
  \label{fig:SALT}
\end{figure}

The RSS spectra are presented in Fig.\,\ref{fig:SALT}. 
Both sources show a single-peaked \Halpha emission line, which dominates the spectrum and is suggestive for a low to intermediate inclination angle of the Be disc.
Also, \Hbeta was seen in emission in both spectra.
The measured equivalent width (EW) and full width at half maximum (FWHM) of the \Halpha lines seen from \asrc and \csrc are summarised in Table\,\ref{tab:saltobs}. In particular, the very strong \Halpha line measured for \csrc indicates a large circum-stellar disc \citep{2006ApJ...651L..53G} and a long orbital period \citep{1997A&A...322..193R}.
Together with \bsrc, which was identified as Be star previously \citep{2002A&A...385..517N,1994PASP..106..843S},
this also confirms \asrc and \csrc as Be/X-ray binary pulsars.

\begin{table*}
\centering
\caption[]{SALT spectroscopy of \asrc and \csrc.}
\label{tab:saltobs}
\begin{tabular}{llrcr}
\hline\hline\noalign{\smallskip}
\multicolumn{1}{c}{Source} &
\multicolumn{1}{c}{Observation} &
\multicolumn{1}{c}{Exposure} &
\multicolumn{1}{c}{EW (\Halpha)} &
\multicolumn{1}{c}{FWHM (\Halpha)} \\
\multicolumn{1}{c}{name} &
\multicolumn{1}{c}{start (UTC)} &
\multicolumn{1}{c}{(s)} &
\multicolumn{1}{c}{(\AA)} &
\multicolumn{1}{c}{(\AA)} \\
\noalign{\smallskip}\hline\noalign{\smallskip}
\asrc & 2020-03-21 18:17:52 &  900 & -10.42 $\pm$ 0.50 & 13.27 $\pm$ 0.10 \\
\noalign{\smallskip}
\csrc & 2021-12-14 20:09:42 & 1600 & -43.9 $\pm$ 1.4   &  9.70 $\pm$ 0.23 \\
\noalign{\smallskip}\hline
\end{tabular}
\end{table*}

\subsection{OGLE monitoring}
\label{sec:ogle}

\begin{table*}
\centering
\caption[]{OGLE monitoring of \asrc, \bsrc, and \csrc.}
\label{tab:ogleobs}
\begin{tabular}{lllll}
\hline\hline\noalign{\smallskip}
\multicolumn{1}{c}{Source} &
\multicolumn{2}{c}{OGLE III} &
\multicolumn{2}{c}{OGLE IV} \\
\multicolumn{1}{c}{name} &
\multicolumn{1}{c}{I-band} &
\multicolumn{1}{c}{V-band} &
\multicolumn{1}{c}{I-band} &
\multicolumn{1}{c}{V-band} \\
\noalign{\smallskip}\hline\noalign{\smallskip}
\asrc &       --        &         --       & LMC512.20.12    & LMC512.20.v.2  \\
\bsrc & LMC129.2.19005  & LMC129.2.v.22055 & LMC508.31.16128 & LMC508.31.v.22929 \\
\csrc & LMC142.3.163    &  LMC142.3.v.52   & LMC531.23.22017 & LMC531.23.v.27359 \\
\noalign{\smallskip}\hline
\end{tabular}
\end{table*}

\begin{figure}[h!]
\centering
 \resizebox{\hsize}{!}{\includegraphics{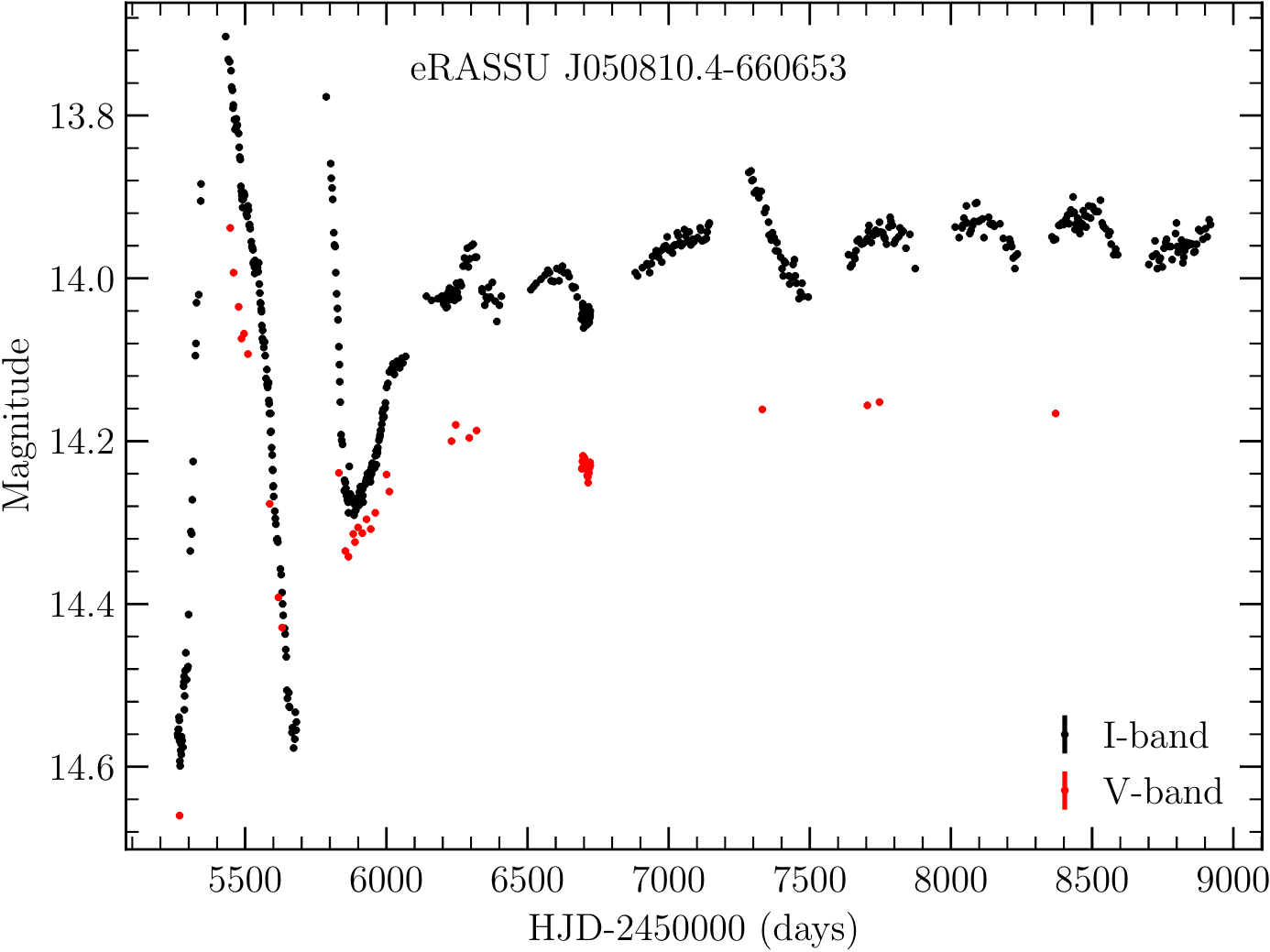}}
  \caption{
    OGLE-IV I- (black) and V-band (red) light curve of \asrc between 2010 March 5 and 2020 March 11. 
  }
  \label{fig:OGLEIa}
\end{figure}

\begin{figure*}
\centering
 \resizebox{\hsize}{!}{\includegraphics{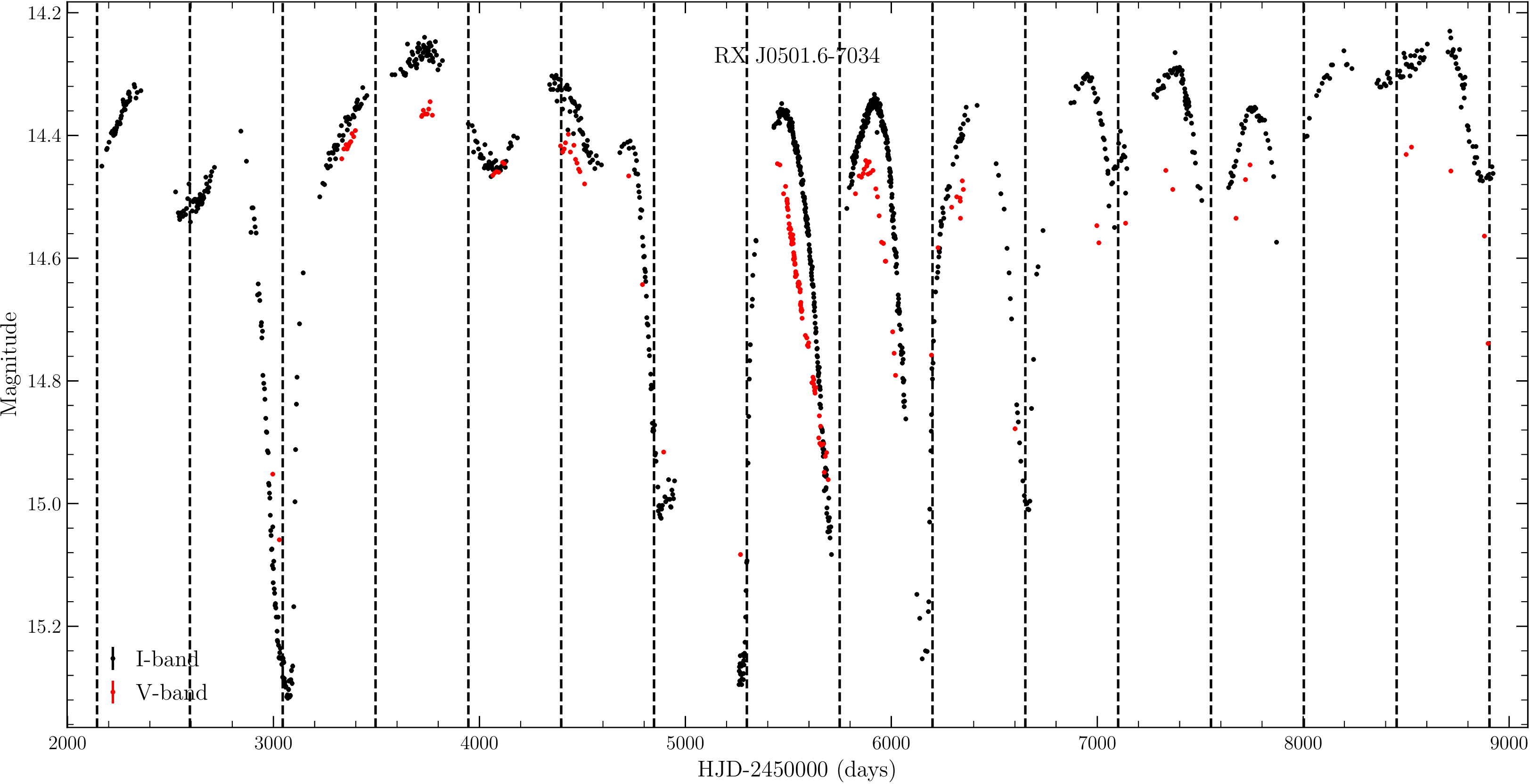}}\\ \vspace{2mm}
 \resizebox{\hsize}{!}{\includegraphics{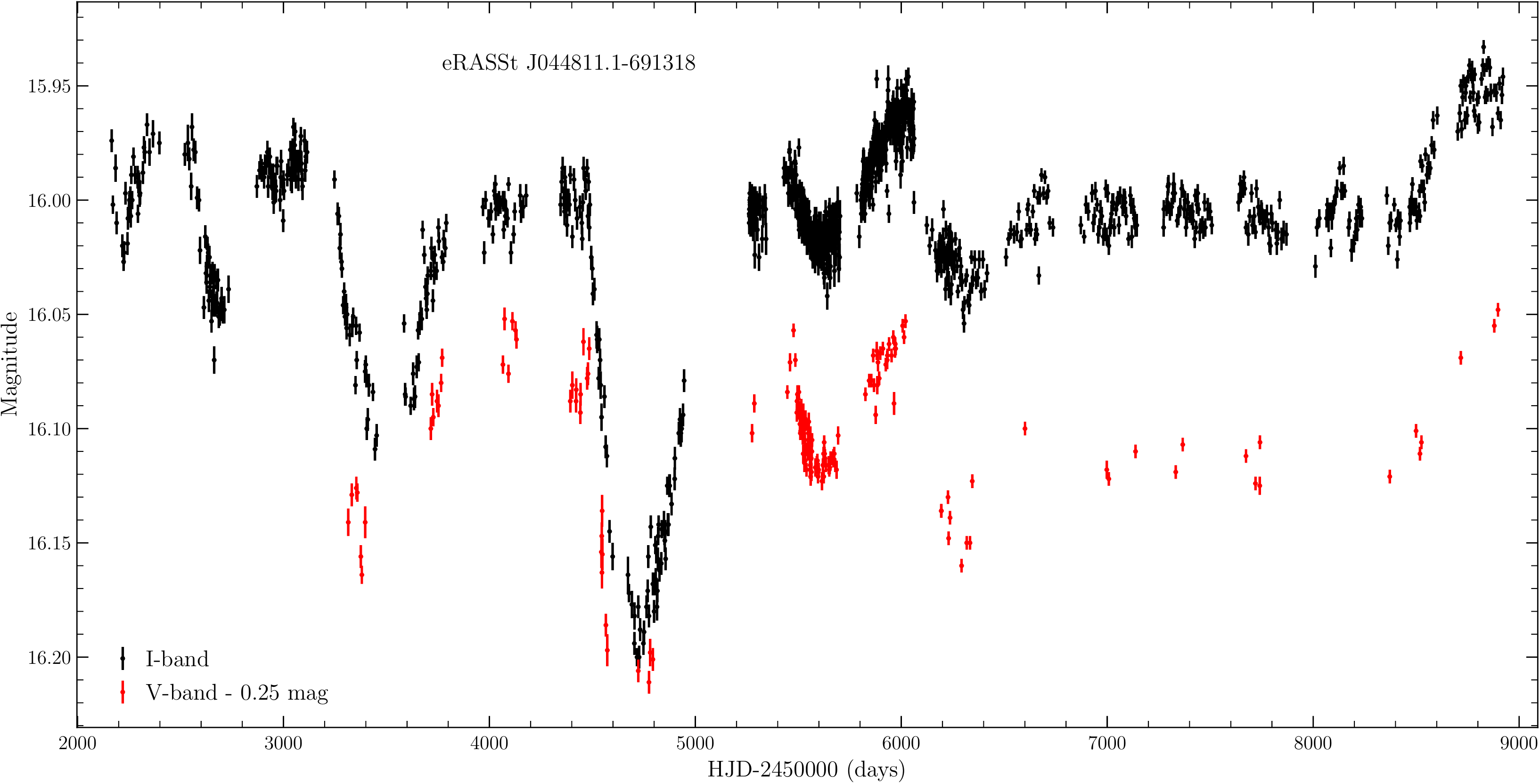}}
  \caption{
    I-band light curves (black) of \bsrc (top; between 2001 September 14 and 2020 March 13) and \csrc (bottom; 2001 September 13 to 2020 March 13) from OGLE phases III and IV. V-band measurements are available for both systems and shown in red. The vertical dashed lines in the top panel are separated by 450.7\,days, the period indicated by the LS periodogram analysis. 
  }
  \label{fig:OGLEIbc}
\end{figure*}

The optical counterparts of the three new Be/X-ray binary pulsars were monitored regularly as part of the OGLE project.
Images were taken in the I- and V-band with the photometric magnitudes calibrated to the standard VI system. 
For all our three targets, their I- and V-band light curves are available, for \asrc during phase IV of the OGLE project 
(ten years of monitoring; see Fig.\,\ref{fig:OGLEIa}) and for \bsrc and \csrc during OGLE phases III and IV 
(in total 19.5 years, Fig.\,\ref{fig:OGLEIbc}). 
Table\,\ref{tab:ogleobs} lists the OGLE IDs of the optical counterparts.

\begin{figure}[h!]
\centering
 \resizebox{0.98\hsize}{!}{\includegraphics{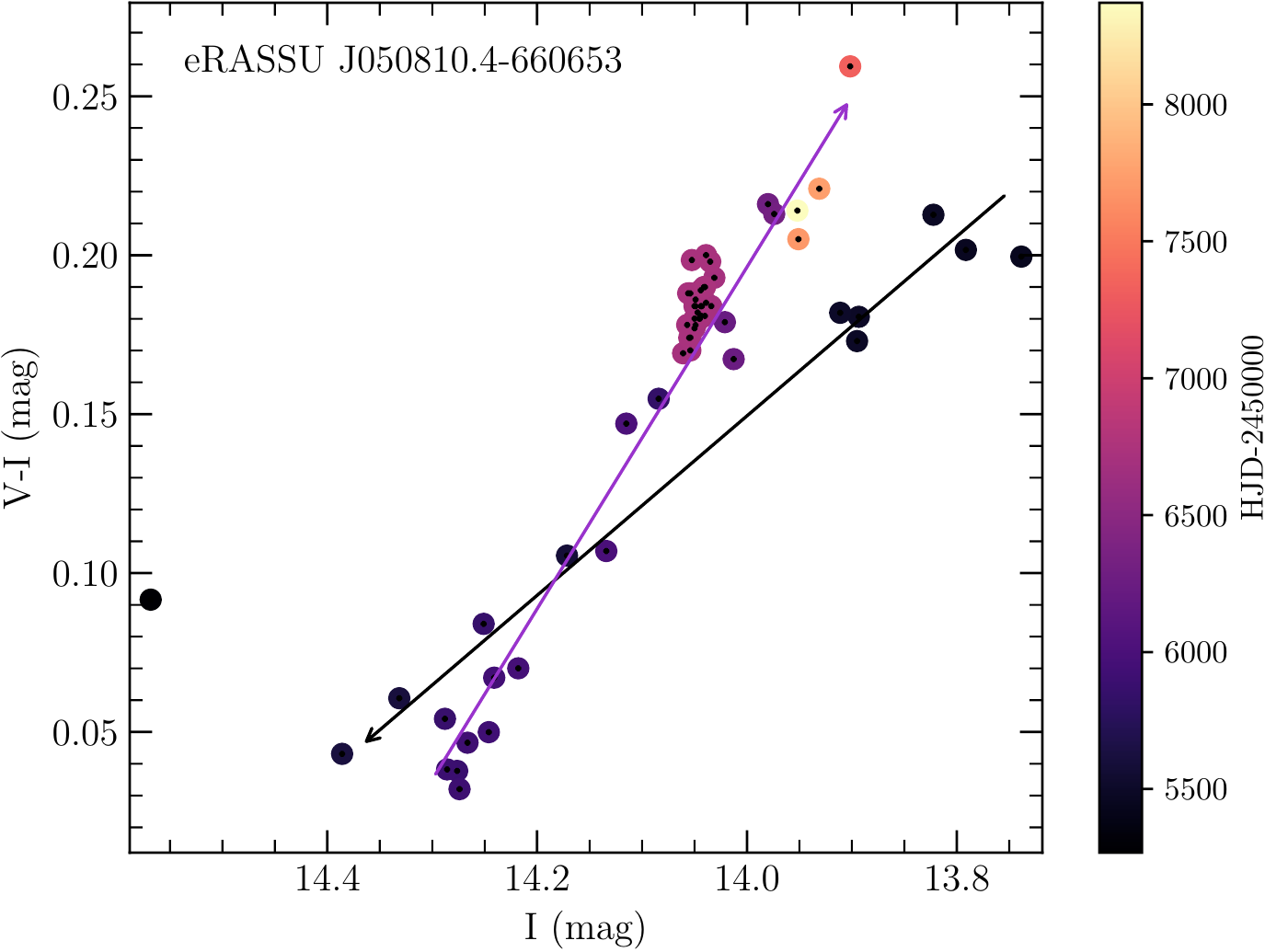}}
 \resizebox{0.98\hsize}{!}{\includegraphics{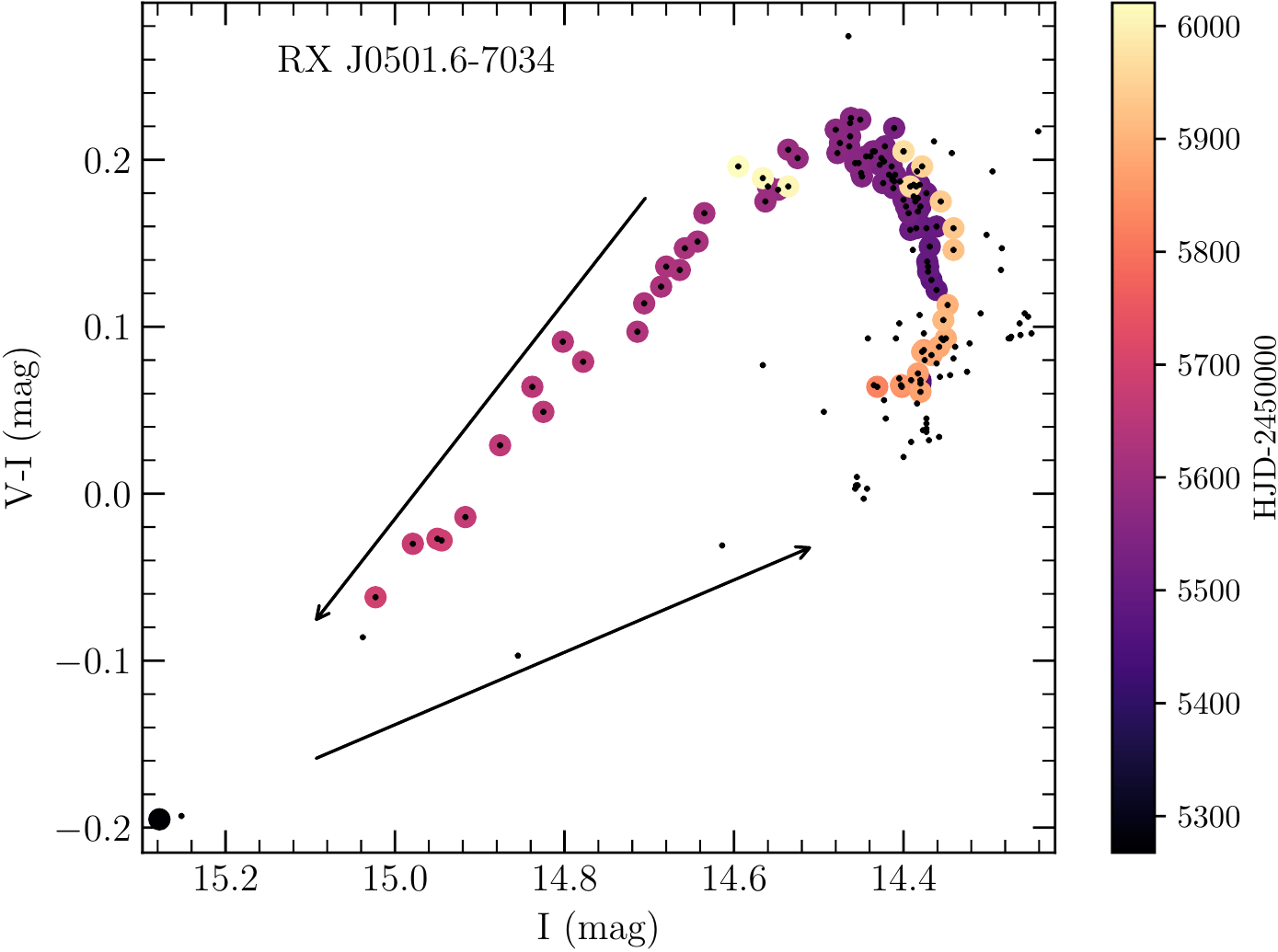}}
 \resizebox{0.98\hsize}{!}{\includegraphics{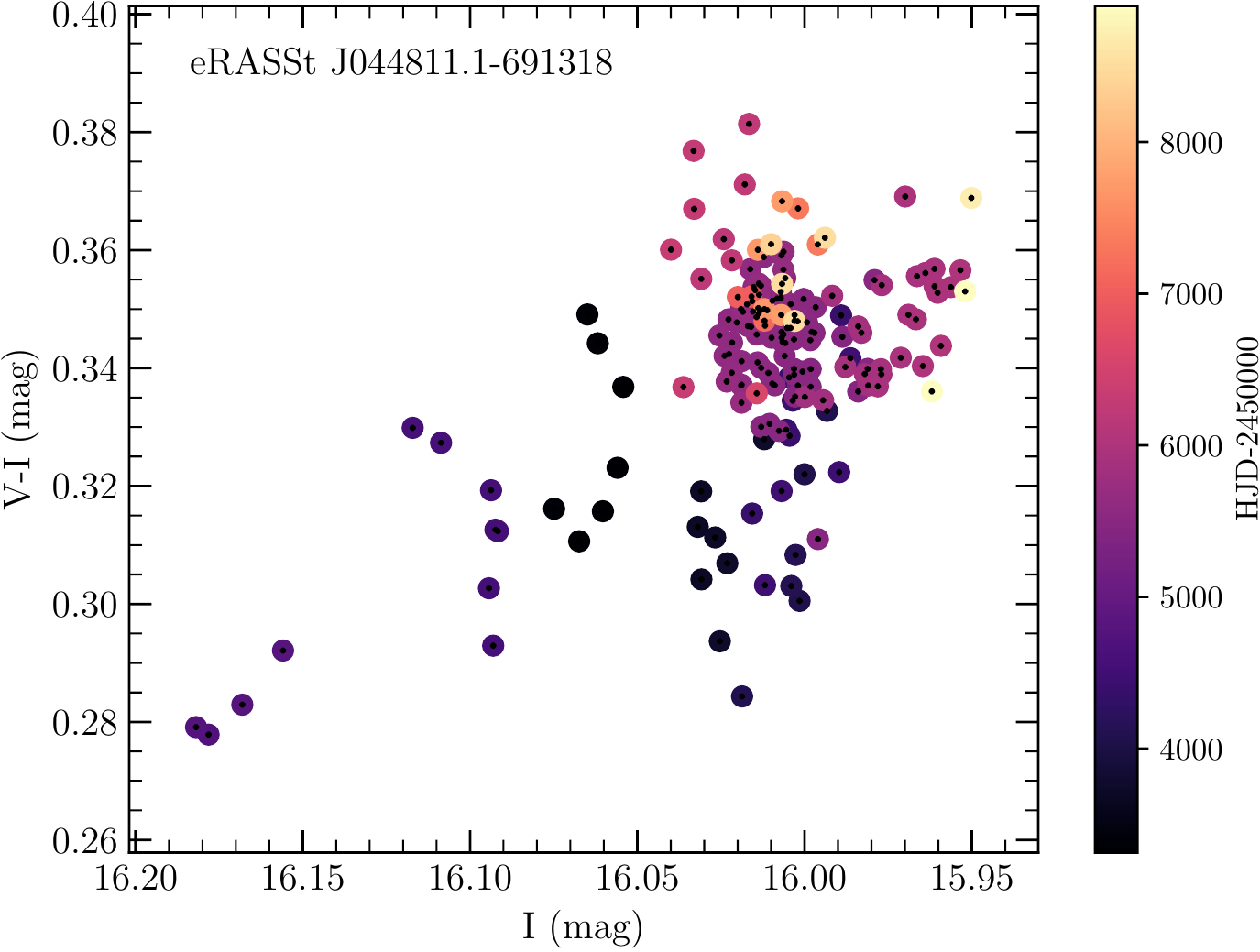}}
  \caption{
    OGLE V-I colour index as a function of I for \asrc ({\it top}), \bsrc ({\it middle}), and \csrc ({\it bottom}). 
    The I values are interpolated to the times of the V-band observations.
    Due to the fast brightness changes seen from \bsrc the interpolated I values are only used when an I measurement exists within five days of the V-band observation. 
  }
  \label{fig:OGLEVI}
\end{figure}

The OGLE light curves of \asrc show large oscillations, initially by nearly one magnitude in I. 
The time between the first two deepest minima is $\sim$420 days.
The decay in brightness after the second maximum seems to have stopped half way 
and the behaviour changed to much smaller brightness variations of $\sim$0.1\,mag in I.
These two phases in the light curves can also be identified in the V-I colour index, 
which is plotted in the top panel of Fig.\,\ref{fig:OGLEVI}. 
During the initial phase the colour index decreases, that is, the system gets bluer when fading in I. 
During the second phase, the system brightens again and returns to a redder colour.
However, it did not reach the maximum brightness again and the colour ended up even redder 
than at the beginning of the light curve. 
In other words, the slopes of the paths in the V-I versus I plane were different during the two phases. 


----------

A similar behaviour is seen in the OGLE light curves of \bsrc. 
Large brightness variations by up to $\sim$1\,mag in the I band occurred several times with 
six deep minima during the OGLE monitoring period. 
The timescale between the deep minima is $\sim$450 days, also very similar to \asrc.
The V-I colour index during the steep brightness declines became also bluer and returned to red 
during brightening in a closed loop as shown in Fig.\,\ref{fig:OGLEVI} (middle panel).

The optical counterpart of \csrc is fainter and the largest variation was a 0.22\,mag brightness 
decrease into a minimum with following recovery, 
which, unfortunately, was not fully covered by the OGLE observations. 
The V-I colour index does not follow such well-defined paths as in the other two systems, 
but \csrc is also redder when brighter (Fig.\,\ref{fig:OGLEVI}, bottom panel).

\begin{figure}
\centering
 \resizebox{0.9\hsize}{!}{\includegraphics{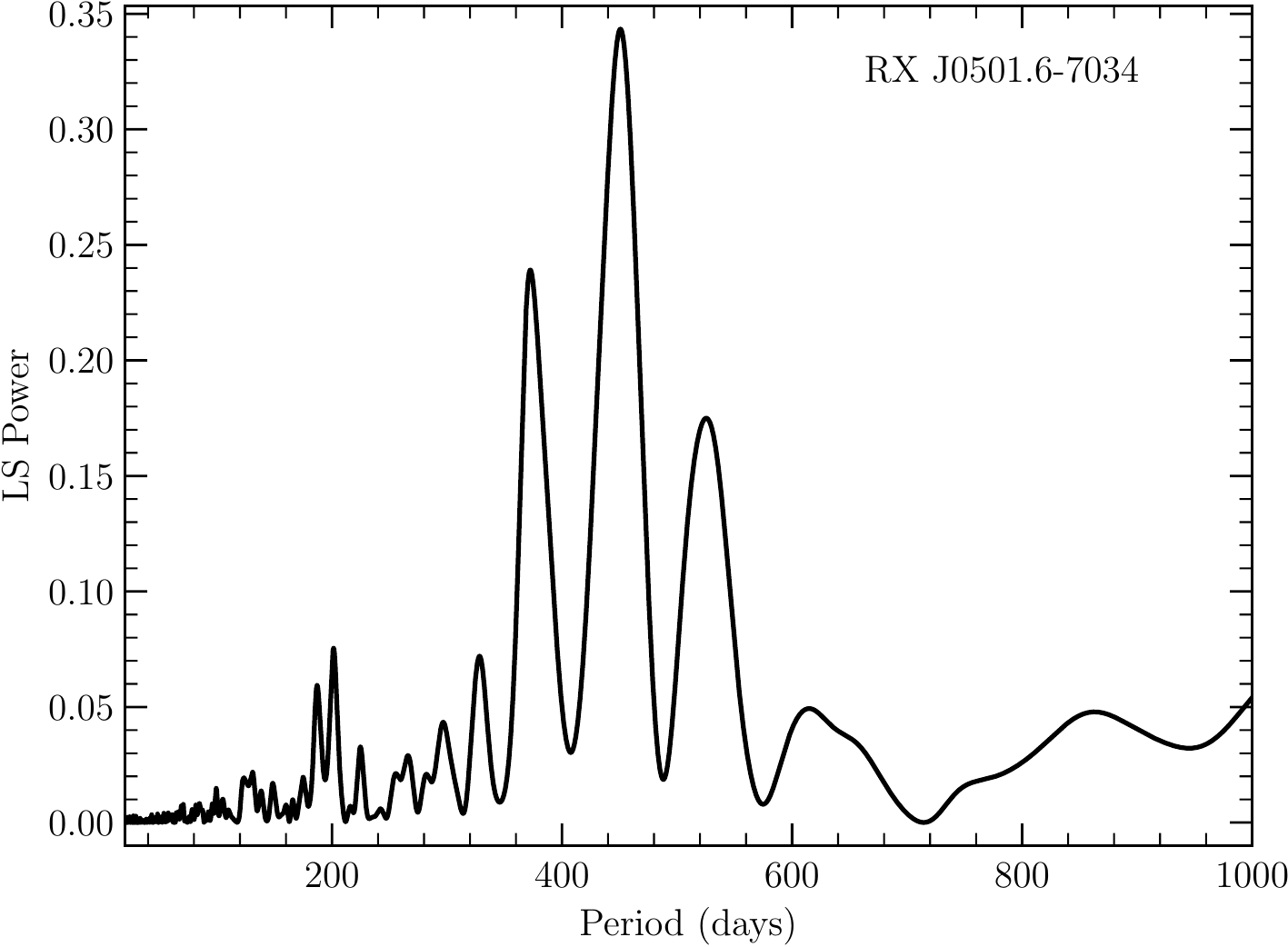}}
  \caption{
    Lomb-Scargle periodogram of the OGLE I-band light curve of \bsrc. The strongest signal is found at a period of 450.7 days.
  }
  \label{fig:OGLEILS}
\end{figure}

The strong variations in the OGLE I-band light curve of \bsrc (Fig.\,\ref{fig:OGLEIbc}, top) follow a regular pattern. 
We conducted a Lomb-Scargle (LS) periodogram analysis \citep{1976Ap&SS..39..447L,1982ApJ...263..835S} in the range of 20--1000\,days to check for a periodic behaviour. 
The strongest peak in the LS periodogram (Fig.\,\ref{fig:OGLEILS}) is indeed found at 450.7\,days. However, the deep dips in the light curves do not occur strictly periodic: some are shifted to earlier phases and some to later ones. 

\section{Discussion}
\label{sec:discussion}

We analysed X-ray and optical data of three HMXB candidates in the LMC. Two of them, \asrc and \csrc, were discovered as new hard X-ray transients in \ero data, while \bsrc was known from observations with the \ein observatory and \ROSAT as candidate for a Be/X-ray binary \citep{1981ApJ...248..925L,1994PASP..106..843S}. The arc-second accuracy of the \xmm/EPIC positions allowed us to uniquely identify the optical counterparts of the new transients and confirm the association of the suggested counterpart SV*\,HV\,2289 with \bsrc.
Their X-ray spectral and temporal properties, together with the long-term behaviour of optical brightness and colours of the counterparts, and, lastly,  the existence of \Halpha emission, confirm all three objects as Be/X-ray binaries.

\begin{figure}
\centering
 \resizebox{0.9\hsize}{!}{\includegraphics{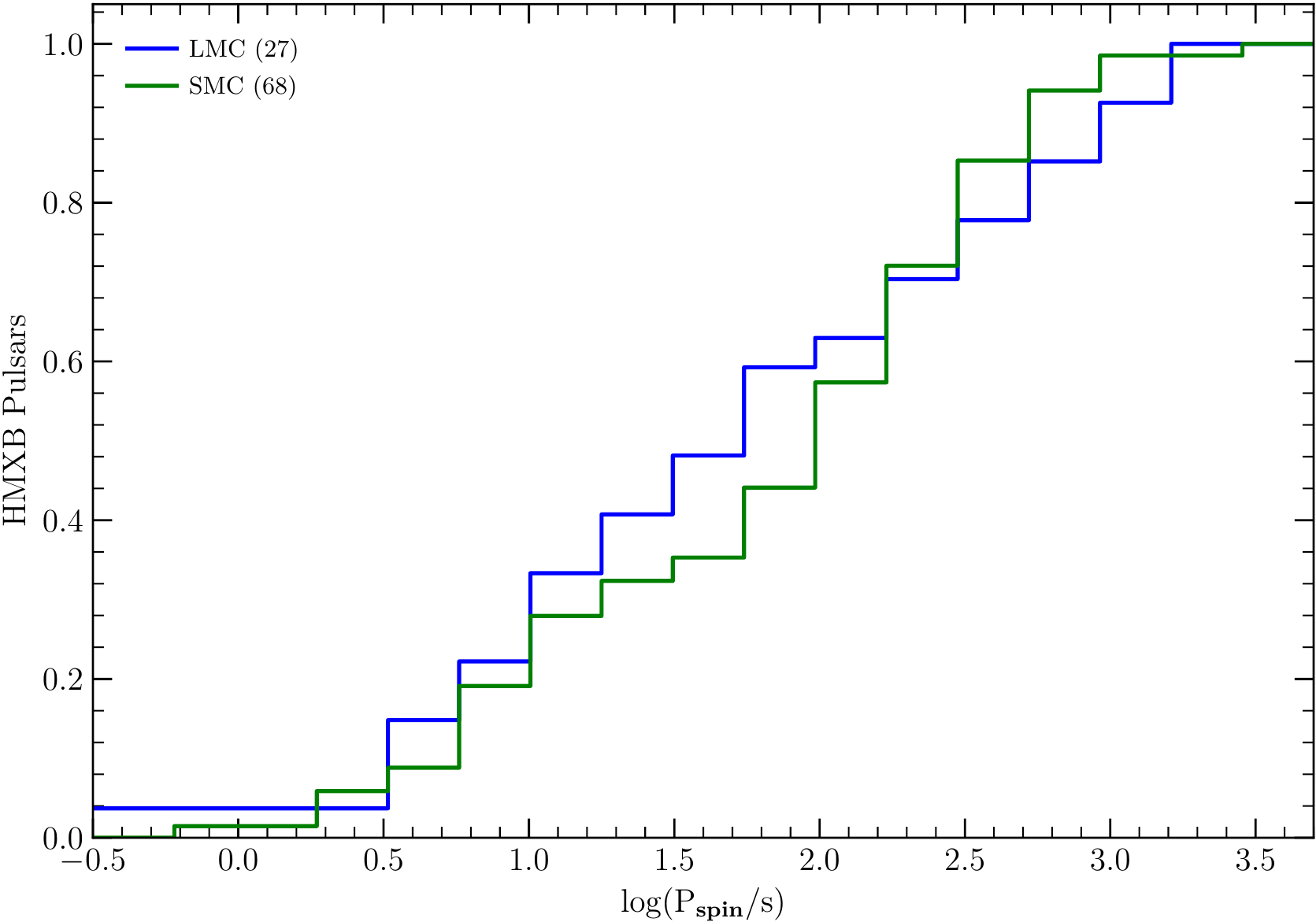}}
  \caption{
    Normalised cumulative spin period distribution of HMXB pulsars in the LMC and SMC.
  }
  \label{fig:MCpulsars}
\end{figure}

In addition, the analysis of the EPIC light curves revealed X-ray pulsations with periods of 40.6\,s, 17.3\,s, and 783.8\,s, for \asrc, \bsrc, and \csrc, respectively. The spin period distribution in the Small Magellanic Cloud (SMC) shows a remarkable bimodality with peaks at $\sim$10\,s and $\sim$200\,s and a minimum at $\sim$25--40\,s 
\citep{2016A&A...586A..81H,2011Natur.479..372K}.
Although the number of known HMXB pulsars in the LMC \citep[27 including the new discoveries from this work;][who had already included  \asrc]{2022A&A...662A..22H} is still significantly lower than in the SMC (68), such a minimum is not obvious for the LMC. 
Figure\,\ref{fig:MCpulsars} compares the cumulative spin period distributions of the known HMXBs in the Magellanic Clouds. The relative abundance of pulsars with spin periods between $\sim$10\,s and $\sim$100\,s is higher in the LMC. However, a statistical two-sample Kolmogorov-Smirnov test reveals only a difference of the cumulative distributions at the $\sim$1\,$\sigma$ level and more LMC systems are required to confirm the difference.
It remains unclear if any of the suggested theories to explain the bimodality seen from the SMC by different types of supernovae \citep{2011Natur.479..372K} or different accretion modes \citep{2014ApJ...786..128C} can also explain a smoother period distribution as possibly seen in the LMC.

To investigate the long-term X-ray light curve of the three BeXRBs we used the HIgh-energy LIght curve GeneraTor \citep[HILIGT;][]{2022A&C....3800529K,2022A&C....3800531S}\footnote{\url{http://xmmuls.esac.esa.int/upperlimitserver/}} to search for serendipitous observations with \xmm and \swift. From the available spectral models to convert count rates to fluxes we used a power law with photon index 1.0 and an absorption column density of \ohcm{21}. We converted the HILIGT 0.2--12\,keV fluxes to the 0.2--10\,keV band by re-normalising them to the values obtained from our spectral fits to the \xmm spectra.
The tool provides upper limits (2$\sigma$) when the source was not detected. We excluded \ROSAT values due to its limited energy band. In the following, we further discuss our results from the sources individually.

\subsection{\asrc}

\begin{figure}
\centering
 \resizebox{0.9\hsize}{!}{\includegraphics{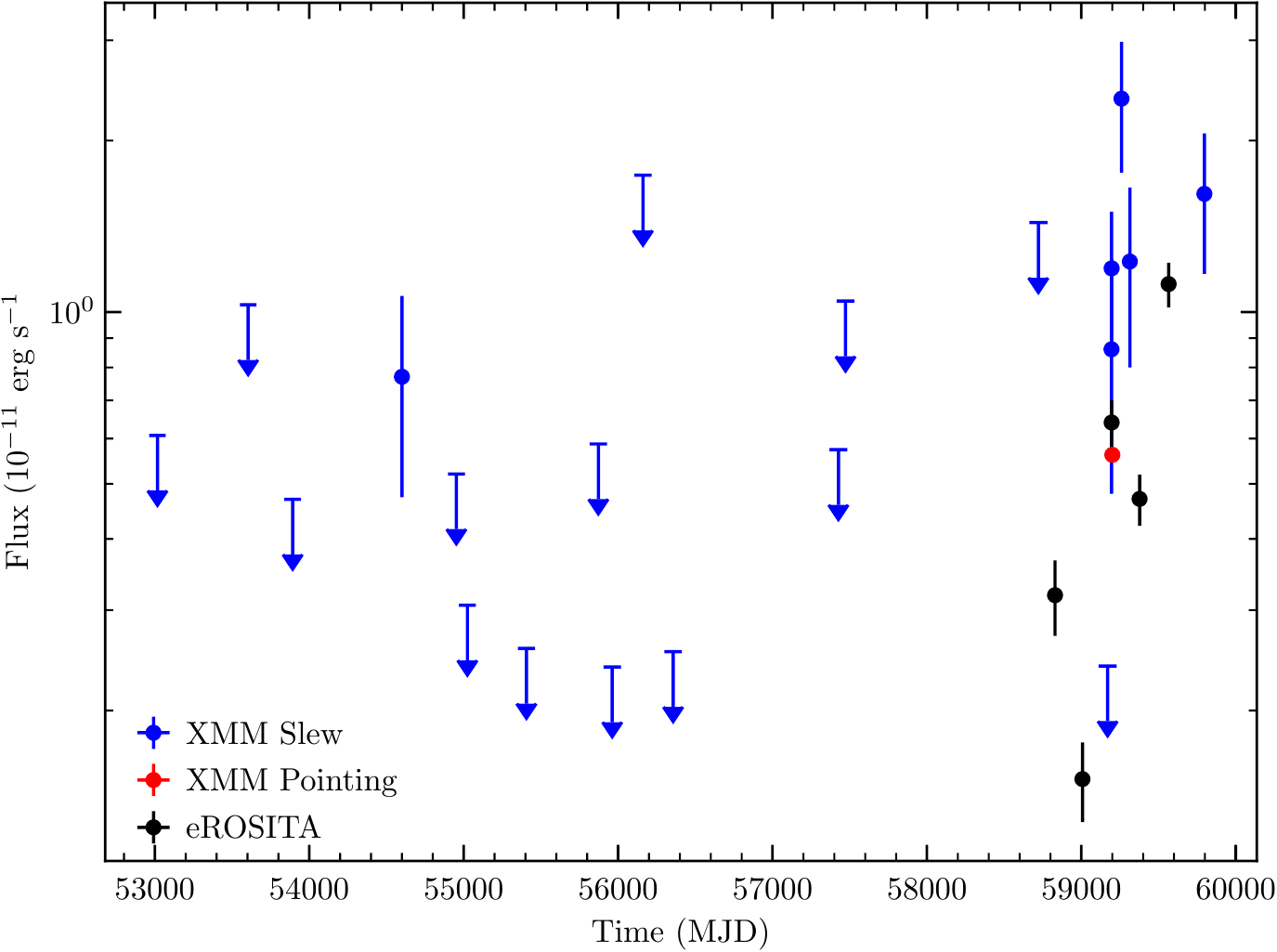}}
  \caption{
    Long-term X-ray light curve of \asrc for the 0.2--10\,keV band. Our \xmm ToO observation is marked in red.
  }
  \label{fig:asrc_HILIGT}
\end{figure}

While most HMXBs in the north of the LMC are located in or near the well-observed supergiant shell LMC\,4, this new BeXRB was found further west between the supergiant shells LMC\,1 and LMC\,5 \citep{1980MNRAS.192..365M}.
The position of \asrc was not covered by earlier \xmm pointed observations, nor by \swift.
HILIGT lists 20 \xmm slews across the position of \asrc, which resulted in six detections, four of these close in time to the ToO observation. During one of them (on 2021 February 14) the highest flux was measured with 2.5\ergcm{-11} while the lowest flux level was inferred from the eRASS1-eRASS2 scans (see Fig.\,\ref{fig:asrc_HILIGT}), resulting in a flux ratio of 15.6$^{+7.8}_{-5.6}$. 
This is typical for persistent low-luminosity BeXRBs that do not exhibit large outbursts \citep{2016A&A...586A..81H} and this behaviour is most likely explained by a wide and nearly circular neutron star orbit \citep{2013MmSAI..84..626L}. 
Several such systems were recently discovered in the LMC \citep{2022A&A...662A..22H,2023A&A...669A..30M}.

\subsection{\bsrc}

\begin{figure}[h!]
\centering
 \resizebox{0.9\hsize}{!}{\includegraphics{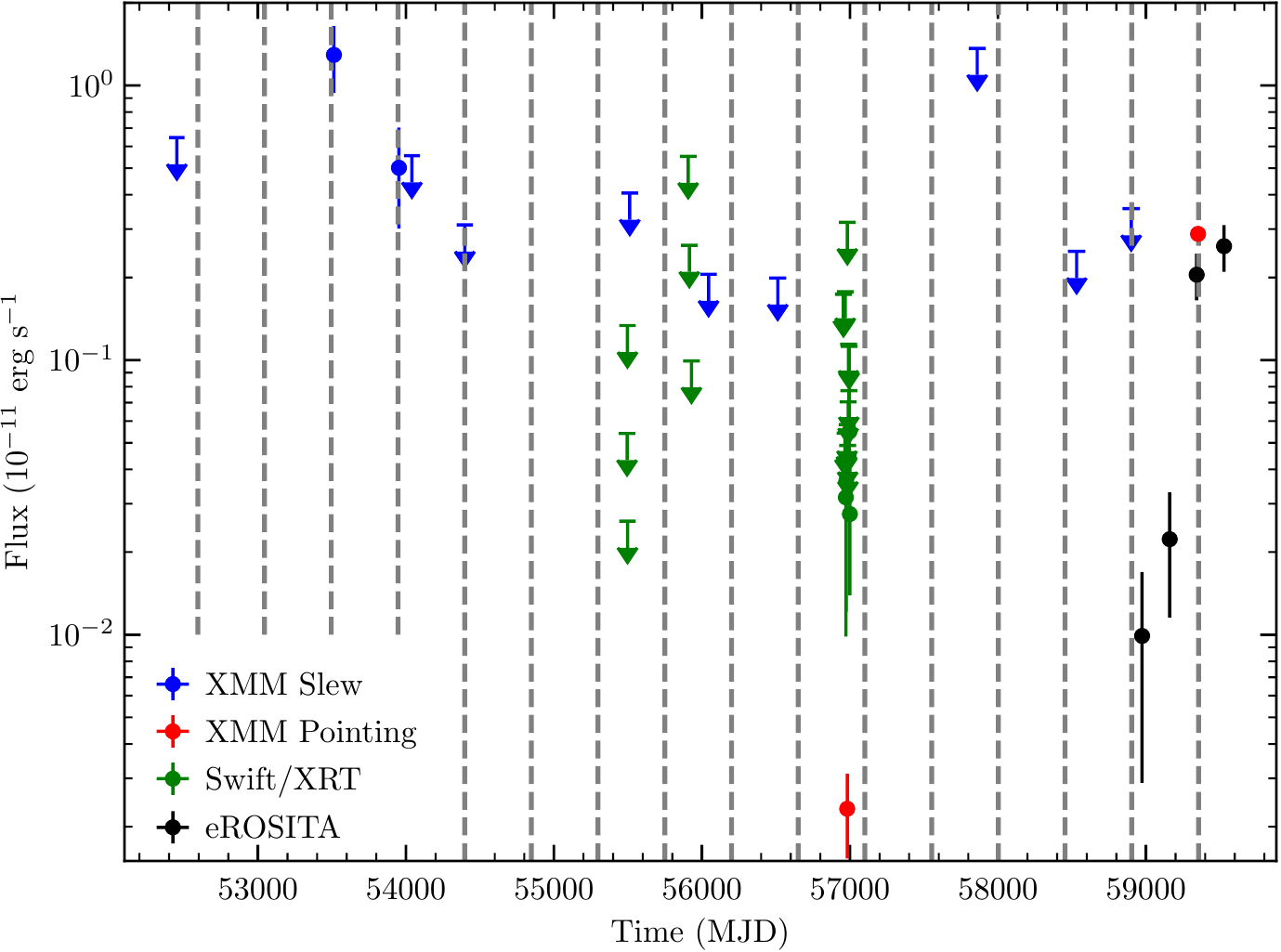}}
  \caption{
    Long-term X-ray (0.2--10\,keV) light curve of \bsrc. The last red data point marks our \xmm ToO observation. The vertical dashed lines mark the times near expected optical minima as seen in Fig.\,\ref{fig:OGLEIbc} (top panel).
  }
  \label{fig:CAL9_HILIGT}
\end{figure}

\bsrc (CAL\,9) was discovered with the \ein observatory with a 0.15--4.5\,keV luminosity of $\sim$4\ergs{35}. Although this estimate is only accurate to a factor of two (due to a model-dependent count rate conversion), it is within the range observed during the \ero and our \xmm ToO observations.
Using HILIGT, we found 11 \xmm slews across \bsrc. During a  slew on 2005 May 23, the highest flux was recorded with 1.3\ergcm{-11}, while the lowest flux was measured at 2.3\ergcm{-14} from an \xmm pointed observation on 2014 November 20 (see Fig.\,\ref{fig:CAL9_HILIGT}), resulting in a flux ratio of 550$^{+520}_{-250}$. 
The position of \bsrc was also covered by 23 \swift/XRT observations. Five detections and 18 upper limits for the flux were all between the minimum and maximum flux seen by \xmm.
The long-term X-ray light curve demonstrates \bsrc as a highly variable BeXRB. Together with the relatively short spin period of the neutron star, the high variability of at least a factor of 300 is in line with the anti-correlation between these two measured variables found by \citet{2016A&A...586A..81H} from BeXRB pulsars in the SMC.

In the OGLE light curves of \bsrc (Fig.\,\ref{fig:OGLEIbc}, top), six extremely deep dips were observed in which the system faded by about one magnitude, while other variations had typical amplitudes of 0.2\,mag. 
These brightness changes repeat nearly periodic every $\sim$451\,days.
However, the sharp deep dips do not occur always strictly at the same phases and shift slightly with time. 
Such a quasi-periodic behaviour suggests a binary orbital period of about 451\,days, with additional effects originating in the interaction of the neutron star with the circum-stellar disc.
Such behaviour was also observed from other BeXRBs in the Magellanic Clouds and could be caused by disc truncation, disc precession, or misalignment between the disc and the orbital plane \citep[][]{2011MNRAS.416.2827M,2012ASPC..464..177O,2021MNRAS.503.6187T}. 

Unfortunately, no X-ray measurements are available during the six deep optical dips. 
However, the highest X-ray fluxes were detected at phases when dips were expected, but did not occur. 
In particular, during the first two \xmm slew detections the system stayed optically bright. 
On the other hand, lowest flux measurements are found between dip phases. 
This suggests the dips occur at the periastron passage of the neutron star, which is also likely to cause the dips. 
The additional long-term effects, as described in the previous paragraph, are then likely to be responsible for the appearance and depth of the dips in the optical light curves. 

A similar hysteresis behaviour with closed loops as shown by \bsrc in the colour-magnitude diagram (Fig.\,\ref{fig:OGLEVI}) was seen from XMMU\,J010331.7$-$730144 \citep{2020MNRAS.496.3615M}. 
The authors discuss periodic mass loss from the Be star to be responsible for the quasi-periodicity of the outbursts and also for the colour-magnitude hysteresis seen from XMMU\,J010331.7$-$730144. 
In their picture, the optically thick inner part of a truncated disc (blue) is removed faster than the outer part (optically thinner, redder) after the mass outflow from the Be star has stopped. The authors argue that the long periodicity of $\sim$1200\,days from XMMU\,J010331.7$-$730144 is unusual for an orbital period, particularly given the relatively small EW of the \Halpha line, which suggests a short orbital period according to the EW - orbital period correlation seen from BeXRBs \citep{1997A&A...322..193R}. 
However, XMMU\,J010331.7$-$730144 and \bsrc also show differences in their OGLE light curves. XMMU\,J010331.7$-$730144 is characterised by extreme outbursts with amplitudes of $\sim$1\,mag in the I-band, which occur quasi-periodically every $\sim$1200\,days, while \bsrc shows 
dips with brightness decreases by a similar amount and on a shorter timescale of $\sim$451\,days. The shorter period found for \bsrc together with a larger \Halpha EW \citep[-10.4\,\AA\ measured in February 2001,][]{2012A&A...539A.114R} may favour the orbital period scenario. In that case the neutron star truncates the outer disc when approaching periastron. This leads to decreasing disc brightness and bluer colour, as indicated by the upper arrow in the middle panel of Fig.\,\ref{fig:OGLEVI}. After periastron passage, the neutron star departs from the disc, which is then  replenished, getting brighter and redder, following the path indicated by the lower arrow.

\subsection{\csrc}

This new BeXRB long-period pulsar (784\,s) is located at the rim of the supergiant shell LMC\,7 in the west of the LMC, where two others had been found: 
XMMU\,J045736.9$-$692727 \citep[318\,s, ][]{2022A&A...662A..22H} and Swift\,J045106.8$-$694803 \citep[187\,s, ][]{2013MNRAS.428.3607K}.  In particular,
\csrc is among the X-ray faintest BeXRBs in the Magellanic Clouds. From 13 \xmm slew observations, only the upper limits of typically a few \oergcm{-12} have been inferred, similar or above the detections with \ero and \xmm (ToO). HILIGT lists two deep archival \xmm  observations on 2017 September 28 with 0.2$-$10\,keV flux of (1.4$\pm$0.6)\ergcm{-14} and 2019 October 15 with (2.0$\pm$0.6)\ergcm{-14}. With the maximum observed flux during eRASS4, the variability is a factor of $\sim$100, but given the large flux uncertainties, it is not very well constrained (100$^{+210}_{-70}$). Nevertheless, this variability is still within the range observed from BeXRB pulsars with long spin periods of nearly 1000\,s \citep{2016A&A...586A..81H}. 

\section{Conclusions}
\label{sec:conclusion}
The discovery of two new Be/X-ray binary systems in \ero data and the detection of pulsations from \bsrc increase the number of Be/X-ray binary pulsars in the LMC to 27. Their spin-period distribution does not show the bimodal characteristics of the SMC pulsars, but more LMC systems are still needed to test if this difference is significant.

\begin{acknowledgements}
This work is based on data from \ero, the soft X-ray instrument aboard \srg, a joint Russian-German science mission supported by the Russian Space Agency (Roskosmos), in the interests of the Russian Academy of Sciences represented by its Space Research Institute (IKI), and the Deutsches Zentrum f{\"u}r Luft- und Raumfahrt (DLR). The \srg spacecraft was built by Lavochkin Association (NPOL) and its subcontractors, and is operated by NPOL with support from the Max Planck Institute for Extraterrestrial Physics (MPE).
The development and construction of the \ero X-ray instrument was led by MPE, with contributions from the Dr. Karl Remeis Observatory Bamberg \& ECAP (FAU Erlangen-N{\"u}rnberg), the University of Hamburg Observatory, the Leibniz Institute for Astrophysics Potsdam (AIP), and the Institute for Astronomy and Astrophysics of the University of T{\"u}bingen, with the support of DLR and the Max Planck Society. The Argelander Institute for Astronomy of the University of Bonn and the Ludwig Maximilians Universit{\"a}t Munich also participated in the science preparation for \ero.
The \ero data shown here were processed using the \eSASS/\nrta software system developed by the German \ero consortium.
This work used observations obtained with \xmm, an ESA science mission with instruments and contributions directly funded by ESA Member States and NASA. The \xmm project is supported by the DLR and the Max Planck Society. 
This research has made use of the VizieR catalogue access tool, CDS,
Strasbourg, France. The original description of the VizieR service was published in A\&AS 143, 23.
This work has made use of data from the European Space Agency (ESA) mission
{\it Gaia} (\url{https://www.cosmos.esa.int/gaia}), processed by the {\it Gaia}
Data Processing and Analysis Consortium (DPAC,
\url{https://www.cosmos.esa.int/web/gaia/dpac/consortium}). Funding for the DPAC
has been provided by national institutions, in particular the institutions
participating in the {\it Gaia} Multilateral Agreement.
LD acknowledges support from the Bundesministerium f{\"u}r Wirtschaft und Energie through the DLR grant FKZ\,50\,QR\,2102.
\end{acknowledgements}

\bibliographystyle{aa} 
\bibliography{general} 

\begin{thebibliography}{57}
\expandafter\ifx\csname natexlab\endcsname\relax\def\natexlab#1{#1}\fi

\bibitem[{{Arnaud}(1996)}]{1996ASPC..101...17A}
{Arnaud}, K.~A. 1996, in ASP Conf. Ser. 101: Astronomical Data Analysis
  Software and Systems V, 17

\bibitem[{{Brunner} {et~al.}(2022){Brunner}, {Liu}, {Lamer}, {Georgakakis},
  {Merloni}, {Brusa}, {Bulbul}, {Dennerl}, {Friedrich}, {Liu}, {Maitra},
  {Nandra}, {Ramos-Ceja}, {Sanders}, {Stewart}, {Boller}, {Buchner}, {Clerc},
  {Comparat}, {Dwelly}, {Eckert}, {Finoguenov}, {Freyberg}, {Ghirardini},
  {Gueguen}, {Haberl}, {Kreykenbohm}, {Krumpe}, {Osterhage}, {Pacaud},
  {Predehl}, {Reiprich}, {Robrade}, {Salvato}, {Santangelo}, {Schrabback},
  {Schwope}, \& {Wilms}}]{2022A&A...661A...1B}
{Brunner}, H., {Liu}, T., {Lamer}, G., {et~al.} 2022, \aap, 661, A1

\bibitem[{{Buckley} {et~al.}(2006){Buckley}, {Swart}, \&
  {Meiring}}]{Buckley2006}
{Buckley}, D. A.~H., {Swart}, G.~P., \& {Meiring}, J.~G. 2006, in Society of
  Photo-Optical Instrumentation Engineers (SPIE) Conference Series, Vol. 6267,
  Society of Photo-Optical Instrumentation Engineers (SPIE) Conference Series,
  ed. L.~M. {Stepp}, 62670Z

\bibitem[{{Burgh} {et~al.}(2003){Burgh}, {Nordsieck}, {Kobulnicky}, {Williams},
  {O'Donoghue}, {Smith}, \& {Percival}}]{2003SPIE.4841.1463B}
{Burgh}, E.~B., {Nordsieck}, K.~H., {Kobulnicky}, H.~A., {et~al.} 2003, in
  Society of Photo-Optical Instrumentation Engineers (SPIE) Conference Series,
  Vol. 4841, Instrument Design and Performance for Optical/Infrared
  Ground-based Telescopes, ed. M.~{Iye} \& A.~F.~M. {Moorwood}, 1463--1471

\bibitem[{{Cash}(1979)}]{1979ApJ...228..939C}
{Cash}, W. 1979, \apj, 228, 939

\bibitem[{{Cheng} {et~al.}(2014){Cheng}, {Shao}, \& {Li}}]{2014ApJ...786..128C}
{Cheng}, Z.-Q., {Shao}, Y., \& {Li}, X.-D. 2014, \apj, 786, 128

\bibitem[{{Cowley} {et~al.}(1984){Cowley}, {Crampton}, {Hutchings}, {Helfand},
  {Hamilton}, {Thorstensen}, \& {Charles}}]{1984ApJ...286..196C}
{Cowley}, A.~P., {Crampton}, D., {Hutchings}, J.~B., {et~al.} 1984, \apj, 286,
  196

\bibitem[{{Crampton} {et~al.}(1985){Crampton}, {Cowley}, {Thompson}, \&
  {Hutchings}}]{1985AJ.....90...43C}
{Crampton}, D., {Cowley}, A.~P., {Thompson}, I.~B., \& {Hutchings}, J.~B. 1985,
  \aj, 90, 43

\bibitem[{{Dickey} \& {Lockman}(1990)}]{1990ARA&A..28..215D}
{Dickey}, J.~M. \& {Lockman}, F.~J. 1990, \araa, 28, 215

\bibitem[{{Gaia Collaboration} {et~al.}(2021){Gaia Collaboration}, {Brown},
  {Vallenari}, {Prusti}, {de Bruijne}, {Babusiaux}, {Biermann}, {Creevey},
  {Evans}, {Eyer}, {Hutton}, {Jansen}, {Jordi}, {Klioner}, {Lammers},
  {Lindegren}, {Luri}, {Mignard}, {Panem}, {Pourbaix}, {Randich}, {Sartoretti},
  {Soubiran}, {Walton}, {Arenou}, {Bailer-Jones}, {Bastian}, {Cropper},
  {Drimmel}, {Katz}, {Lattanzi}, {van Leeuwen}, {Bakker}, {Cacciari},
  {Casta{\~n}eda}, {De Angeli}, {Ducourant}, {Fabricius}, {Fouesneau},
  {Fr{\'e}mat}, {Guerra}, {Guerrier}, {Guiraud}, {Jean-Antoine Piccolo},
  {Masana}, {Messineo}, {Mowlavi}, {Nicolas}, {Nienartowicz}, {Pailler},
  {Panuzzo}, {Riclet}, {Roux}, {Seabroke}, {Sordo}, {Tanga}, {Th{\'e}venin},
  {Gracia-Abril}, {Portell}, {Teyssier}, {Altmann}, {Andrae}, {Bellas-Velidis},
  {Benson}, {Berthier}, {Blomme}, {Brugaletta}, {Burgess}, {Busso}, {Carry},
  {Cellino}, {Cheek}, {Clementini}, {Damerdji}, {Davidson}, {Delchambre},
  {Dell'Oro}, {Fern{\'a}ndez-Hern{\'a}ndez}, {Galluccio}, {Garc{\'\i}a-Lario},
  {Garcia-Reinaldos}, {Gonz{\'a}lez-N{\'u}{\~n}ez}, {Gosset}, {Haigron},
  {Halbwachs}, {Hambly}, {Harrison}, {Hatzidimitriou}, {Heiter},
  {Hern{\'a}ndez}, {Hestroffer}, {Hodgkin}, {Holl}, {Jan{\ss}en}, {Jevardat de
  Fombelle}, {Jordan}, {Krone-Martins}, {Lanzafame}, {L{\"o}ffler}, {Lorca},
  {Manteiga}, {Marchal}, {Marrese}, {Moitinho}, {Mora}, {Muinonen}, {Osborne},
  {Pancino}, {Pauwels}, {Petit}, {Recio-Blanco}, {Richards}, {Riello},
  {Rimoldini}, {Robin}, {Roegiers}, {Rybizki}, {Sarro}, {Siopis}, {Smith},
  {Sozzetti}, {Ulla}, {Utrilla}, {van Leeuwen}, {van Reeven}, {Abbas}, {Abreu
  Aramburu}, {Accart}, {Aerts}, {Aguado}, {Ajaj}, {Altavilla}, {{\'A}lvarez},
  {{\'A}lvarez Cid-Fuentes}, {Alves}, {Anderson}, {Anglada Varela}, {Antoja},
  {Audard}, {Baines}, {Baker}, {Balaguer-N{\'u}{\~n}ez}, {Balbinot}, {Balog},
  {Barache}, {Barbato}, {Barros}, {Barstow}, {Bartolom{\'e}}, {Bassilana},
  {Bauchet}, {Baudesson-Stella}, {Becciani}, {Bellazzini}, {Bernet}, {Bertone},
  {Bianchi}, {Blanco-Cuaresma}, {Boch}, {Bombrun}, {Bossini}, {Bouquillon},
  {Bragaglia}, {Bramante}, {Breedt}, {Bressan}, {Brouillet}, {Bucciarelli},
  {Burlacu}, {Busonero}, {Butkevich}, {Buzzi}, {Caffau}, {Cancelliere},
  {C{\'a}novas}, {Cantat-Gaudin}, {Carballo}, {Carlucci}, {Carnerero},
  {Carrasco}, {Casamiquela}, {Castellani}, {Castro-Ginard}, {Castro Sampol},
  {Chaoul}, {Charlot}, {Chemin}, {Chiavassa}, {Cioni}, {Comoretto}, {Cooper},
  {Cornez}, {Cowell}, {Crifo}, {Crosta}, {Crowley}, {Dafonte}, {Dapergolas},
  {David}, {David}, {de Laverny}, {De Luise}, {De March}, {De Ridder}, {de
  Souza}, {de Teodoro}, {de Torres}, {del Peloso}, {del Pozo}, {Delbo},
  {Delgado}, {Delgado}, {Delisle}, {Di Matteo}, {Diakite}, {Diener},
  {Distefano}, {Dolding}, {Eappachen}, {Edvardsson}, {Enke}, {Esquej}, {Fabre},
  {Fabrizio}, {Faigler}, {Fedorets}, {Fernique}, {Fienga}, {Figueras},
  {Fouron}, {Fragkoudi}, {Fraile}, {Franke}, {Gai}, {Garabato},
  {Garcia-Gutierrez}, {Garc{\'\i}a-Torres}, {Garofalo}, {Gavras}, {Gerlach},
  {Geyer}, {Giacobbe}, {Gilmore}, {Girona}, {Giuffrida}, {Gomel}, {Gomez},
  {Gonzalez-Santamaria}, {Gonz{\'a}lez-Vidal}, {Granvik},
  {Guti{\'e}rrez-S{\'a}nchez}, {Guy}, {Hauser}, {Haywood}, {Helmi}, {Hidalgo},
  {Hilger}, {H{\l}adczuk}, {Hobbs}, {Holland}, {Huckle}, {Jasniewicz},
  {Jonker}, {Juaristi Campillo}, {Julbe}, {Karbevska}, {Kervella}, {Khanna},
  {Kochoska}, {Kontizas}, {Kordopatis}, {Korn}, {Kostrzewa-Rutkowska},
  {Kruszy{\'n}ska}, {Lambert}, {Lanza}, {Lasne}, {Le Campion}, {Le Fustec},
  {Lebreton}, {Lebzelter}, {Leccia}, {Leclerc}, {Lecoeur-Taibi}, {Liao},
  {Licata}, {Lindstr{\o}m}, {Lister}, {Livanou}, {Lobel}, {Madrero Pardo},
  {Managau}, {Mann}, {Marchant}, {Marconi}, {Marcos Santos}, {Marinoni},
  {Marocco}, {Marshall}, {Martin Polo}, {Mart{\'\i}n-Fleitas}, {Masip},
  {Massari}, {Mastrobuono-Battisti}, {Mazeh}, {McMillan}, {Messina},
  {Michalik}, {Millar}, {Mints}, {Molina}, {Molinaro}, {Moln{\'a}r},
  {Montegriffo}, {Mor}, {Morbidelli}, {Morel}, {Morris}, {Mulone}, {Munoz},
  {Muraveva}, {Murphy}, {Musella}, {Noval}, {Ord{\'e}novic}, {Orr{\`u}},
  {Osinde}, {Pagani}, {Pagano}, {Palaversa}, {Palicio}, {Panahi}, {Pawlak},
  {Pe{\~n}alosa Esteller}, {Penttil{\"a}}, {Piersimoni}, {Pineau}, {Plachy},
  {Plum}, {Poggio}, {Poretti}, {Poujoulet}, {Pr{\v{s}}a}, {Pulone}, {Racero},
  {Ragaini}, {Rainer}, {Raiteri}, {Rambaux}, {Ramos}, {Ramos-Lerate}, {Re
  Fiorentin}, {Regibo}, {Reyl{\'e}}, {Ripepi}, {Riva}, {Rixon}, {Robichon},
  {Robin}, {Roelens}, {Rohrbasser}, {Romero-G{\'o}mez}, {Rowell}, {Royer},
  {Rybicki}, {Sadowski}, {Sagrist{\`a} Sell{\'e}s}, {Sahlmann}, {Salgado},
  {Salguero}, {Samaras}, {Sanchez Gimenez}, {Sanna}, {Santove{\~n}a},
  {Sarasso}, {Schultheis}, {Sciacca}, {Segol}, {Segovia}, {S{\'e}gransan},
  {Semeux}, {Shahaf}, {Siddiqui}, {Siebert}, {Siltala}, {Slezak}, {Smart},
  {Solano}, {Solitro}, {Souami}, {Souchay}, {Spagna}, {Spoto}, {Steele},
  {Steidelm{\"u}ller}, {Stephenson}, {S{\"u}veges}, {Szabados}, {Szegedi-Elek},
  {Taris}, {Tauran}, {Taylor}, {Teixeira}, {Thuillot}, {Tonello}, {Torra},
  {Torra}, {Turon}, {Unger}, {Vaillant}, {van Dillen}, {Vanel}, {Vecchiato},
  {Viala}, {Vicente}, {Voutsinas}, {Weiler}, {Wevers}, {Wyrzykowski}, {Yoldas},
  {Yvard}, {Zhao}, {Zorec}, {Zucker}, {Zurbach}, \&
  {Zwitter}}]{2021A&A...649A...1G}
{Gaia Collaboration}, {Brown}, A.~G.~A., {Vallenari}, A., {et~al.} 2021, \aap,
  649, A1

\bibitem[{{Gaia Collaboration} {et~al.}(2016){Gaia Collaboration}, {Prusti},
  {de Bruijne}, {Brown}, {Vallenari}, {Babusiaux}, {Bailer-Jones}, {Bastian},
  {Biermann}, {Evans}, {Eyer}, {Jansen}, {Jordi}, {Klioner}, {Lammers},
  {Lindegren}, {Luri}, {Mignard}, {Milligan}, {Panem}, {Poinsignon},
  {Pourbaix}, {Randich}, {Sarri}, {Sartoretti}, {Siddiqui}, {Soubiran},
  {Valette}, {van Leeuwen}, {Walton}, {Aerts}, {Arenou}, {Cropper}, {Drimmel},
  {H{\o}g}, {Katz}, {Lattanzi}, {O'Mullane}, {Grebel}, {Holland}, {Huc},
  {Passot}, {Bramante}, {Cacciari}, {Casta{\~n}eda}, {Chaoul}, {Cheek}, {De
  Angeli}, {Fabricius}, {Guerra}, {Hern{\'a}ndez}, {Jean-Antoine-Piccolo},
  {Masana}, {Messineo}, {Mowlavi}, {Nienartowicz}, {Ord{\'o}{\~n}ez-Blanco},
  {Panuzzo}, {Portell}, {Richards}, {Riello}, {Seabroke}, {Tanga},
  {Th{\'e}venin}, {Torra}, {Els}, {Gracia-Abril}, {Comoretto},
  {Garcia-Reinaldos}, {Lock}, {Mercier}, {Altmann}, {Andrae}, {Astraatmadja},
  {Bellas-Velidis}, {Benson}, {Berthier}, {Blomme}, {Busso}, {Carry},
  {Cellino}, {Clementini}, {Cowell}, {Creevey}, {Cuypers}, {Davidson}, {De
  Ridder}, {de Torres}, {Delchambre}, {Dell'Oro}, {Ducourant}, {Fr{\'e}mat},
  {Garc{\'\i}a-Torres}, {Gosset}, {Halbwachs}, {Hambly}, {Harrison}, {Hauser},
  {Hestroffer}, {Hodgkin}, {Huckle}, {Hutton}, {Jasniewicz}, {Jordan},
  {Kontizas}, {Korn}, {Lanzafame}, {Manteiga}, {Moitinho}, {Muinonen},
  {Osinde}, {Pancino}, {Pauwels}, {Petit}, {Recio-Blanco}, {Robin}, {Sarro},
  {Siopis}, {Smith}, {Smith}, {Sozzetti}, {Thuillot}, {van Reeven}, {Viala},
  {Abbas}, {Abreu Aramburu}, {Accart}, {Aguado}, {Allan}, {Allasia},
  {Altavilla}, {{\'A}lvarez}, {Alves}, {Anderson}, {Andrei}, {Anglada Varela},
  {Antiche}, {Antoja}, {Ant{\'o}n}, {Arcay}, {Atzei}, {Ayache}, {Bach},
  {Baker}, {Balaguer-N{\'u}{\~n}ez}, {Barache}, {Barata}, {Barbier}, {Barblan},
  {Baroni}, {Barrado y Navascu{\'e}s}, {Barros}, {Barstow}, {Becciani},
  {Bellazzini}, {Bellei}, {Bello Garc{\'\i}a}, {Belokurov}, {Bendjoya},
  {Berihuete}, {Bianchi}, {Bienaym{\'e}}, {Billebaud}, {Blagorodnova},
  {Blanco-Cuaresma}, {Boch}, {Bombrun}, {Borrachero}, {Bouquillon}, {Bourda},
  {Bouy}, {Bragaglia}, {Breddels}, {Brouillet}, {Br{\"u}semeister},
  {Bucciarelli}, {Budnik}, {Burgess}, {Burgon}, {Burlacu}, {Busonero}, {Buzzi},
  {Caffau}, {Cambras}, {Campbell}, {Cancelliere}, {Cantat-Gaudin}, {Carlucci},
  {Carrasco}, {Castellani}, {Charlot}, {Charnas}, {Charvet}, {Chassat},
  {Chiavassa}, {Clotet}, {Cocozza}, {Collins}, {Collins}, {Costigan}, {Crifo},
  {Cross}, {Crosta}, {Crowley}, {Dafonte}, {Damerdji}, {Dapergolas}, {David},
  {David}, {De Cat}, {de Felice}, {de Laverny}, {De Luise}, {De March}, {de
  Martino}, {de Souza}, {Debosscher}, {del Pozo}, {Delbo}, {Delgado},
  {Delgado}, {di Marco}, {Di Matteo}, {Diakite}, {Distefano}, {Dolding}, {Dos
  Anjos}, {Drazinos}, {Dur{\'a}n}, {Dzigan}, {Ecale}, {Edvardsson}, {Enke},
  {Erdmann}, {Escolar}, {Espina}, {Evans}, {Eynard Bontemps}, {Fabre},
  {Fabrizio}, {Faigler}, {Falc{\~a}o}, {Farr{\`a}s Casas}, {Faye}, {Federici},
  {Fedorets}, {Fern{\'a}ndez-Hern{\'a}ndez}, {Fernique}, {Fienga}, {Figueras},
  {Filippi}, {Findeisen}, {Fonti}, {Fouesneau}, {Fraile}, {Fraser}, {Fuchs},
  {Furnell}, {Gai}, {Galleti}, {Galluccio}, {Garabato}, {Garc{\'\i}a-Sedano},
  {Gar{\'e}}, {Garofalo}, {Garralda}, {Gavras}, {Gerssen}, {Geyer}, {Gilmore},
  {Girona}, {Giuffrida}, {Gomes}, {Gonz{\'a}lez-Marcos},
  {Gonz{\'a}lez-N{\'u}{\~n}ez}, {Gonz{\'a}lez-Vidal}, {Granvik}, {Guerrier},
  {Guillout}, {Guiraud}, {G{\'u}rpide}, {Guti{\'e}rrez-S{\'a}nchez}, {Guy},
  {Haigron}, {Hatzidimitriou}, {Haywood}, {Heiter}, {Helmi}, {Hobbs},
  {Hofmann}, {Holl}, {Holland}, {Hunt}, {Hypki}, {Icardi}, {Irwin}, {Jevardat
  de Fombelle}, {Jofr{\'e}}, {Jonker}, {Jorissen}, {Julbe}, {Karampelas},
  {Kochoska}, {Kohley}, {Kolenberg}, {Kontizas}, {Koposov}, {Kordopatis},
  {Koubsky}, {Kowalczyk}, {Krone-Martins}, {Kudryashova}, {Kull}, {Bachchan},
  {Lacoste-Seris}, {Lanza}, {Lavigne}, {Le Poncin-Lafitte}, {Lebreton},
  {Lebzelter}, {Leccia}, {Leclerc}, {Lecoeur-Taibi}, {Lemaitre}, {Lenhardt},
  {Leroux}, {Liao}, {Licata}, {Lindstr{\o}m}, {Lister}, {Livanou}, {Lobel},
  {L{\"o}ffler}, {L{\'o}pez}, {Lopez-Lozano}, {Lorenz}, {Loureiro},
  {MacDonald}, {Magalh{\~a}es Fernandes}, {Managau}, {Mann}, {Mantelet},
  {Marchal}, {Marchant}, {Marconi}, {Marie}, {Marinoni}, {Marrese},
  {Marschalk{\'o}}, {Marshall}, {Mart{\'\i}n-Fleitas}, {Martino}, {Mary},
  {Matijevi{\v{c}}}, {Mazeh}, {McMillan}, {Messina}, {Mestre}, {Michalik},
  {Millar}, {Miranda}, {Molina}, {Molinaro}, {Molinaro}, {Moln{\'a}r},
  {Moniez}, {Montegriffo}, {Monteiro}, {Mor}, {Mora}, {Morbidelli}, {Morel},
  {Morgenthaler}, {Morley}, {Morris}, {Mulone}, {Muraveva}, {Musella},
  {Narbonne}, {Nelemans}, {Nicastro}, {Noval}, {Ord{\'e}novic},
  {Ordieres-Mer{\'e}}, {Osborne}, {Pagani}, {Pagano}, {Pailler}, {Palacin},
  {Palaversa}, {Parsons}, {Paulsen}, {Pecoraro}, {Pedrosa}, {Pentik{\"a}inen},
  {Pereira}, {Pichon}, {Piersimoni}, {Pineau}, {Plachy}, {Plum}, {Poujoulet},
  {Pr{\v{s}}a}, {Pulone}, {Ragaini}, {Rago}, {Rambaux}, {Ramos-Lerate},
  {Ranalli}, {Rauw}, {Read}, {Regibo}, {Renk}, {Reyl{\'e}}, {Ribeiro},
  {Rimoldini}, {Ripepi}, {Riva}, {Rixon}, {Roelens}, {Romero-G{\'o}mez},
  {Rowell}, {Royer}, {Rudolph}, {Ruiz-Dern}, {Sadowski}, {Sagrist{\`a}
  Sell{\'e}s}, {Sahlmann}, {Salgado}, {Salguero}, {Sarasso}, {Savietto},
  {Schnorhk}, {Schultheis}, {Sciacca}, {Segol}, {Segovia}, {Segransan},
  {Serpell}, {Shih}, {Smareglia}, {Smart}, {Smith}, {Solano}, {Solitro},
  {Sordo}, {Soria Nieto}, {Souchay}, {Spagna}, {Spoto}, {Stampa}, {Steele},
  {Steidelm{\"u}ller}, {Stephenson}, {Stoev}, {Suess}, {S{\"u}veges}, {Surdej},
  {Szabados}, {Szegedi-Elek}, {Tapiador}, {Taris}, {Tauran}, {Taylor},
  {Teixeira}, {Terrett}, {Tingley}, {Trager}, {Turon}, {Ulla}, {Utrilla},
  {Valentini}, {van Elteren}, {Van Hemelryck}, {van Leeuwen}, {Varadi},
  {Vecchiato}, {Veljanoski}, {Via}, {Vicente}, {Vogt}, {Voss}, {Votruba},
  {Voutsinas}, {Walmsley}, {Weiler}, {Weingrill}, {Werner}, {Wevers},
  {Whitehead}, {Wyrzykowski}, {Yoldas}, {{\v{Z}}erjal}, {Zucker}, {Zurbach},
  {Zwitter}, {Alecu}, {Allen}, {Allende Prieto}, {Amorim},
  {Anglada-Escud{\'e}}, {Arsenijevic}, {Azaz}, {Balm}, {Beck}, {Bernstein},
  {Bigot}, {Bijaoui}, {Blasco}, {Bonfigli}, {Bono}, {Boudreault}, {Bressan},
  {Brown}, {Brunet}, {Bunclark}, {Buonanno}, {Butkevich}, {Carret}, {Carrion},
  {Chemin}, {Ch{\'e}reau}, {Corcione}, {Darmigny}, {de Boer}, {de Teodoro}, {de
  Zeeuw}, {Delle Luche}, {Domingues}, {Dubath}, {Fodor}, {Fr{\'e}zouls},
  {Fries}, {Fustes}, {Fyfe}, {Gallardo}, {Gallegos}, {Gardiol}, {Gebran},
  {Gomboc}, {G{\'o}mez}, {Grux}, {Gueguen}, {Heyrovsky}, {Hoar}, {Iannicola},
  {Isasi Parache}, {Janotto}, {Joliet}, {Jonckheere}, {Keil}, {Kim},
  {Klagyivik}, {Klar}, {Knude}, {Kochukhov}, {Kolka}, {Kos}, {Kutka}, {Lainey},
  {LeBouquin}, {Liu}, {Loreggia}, {Makarov}, {Marseille}, {Martayan},
  {Martinez-Rubi}, {Massart}, {Meynadier}, {Mignot}, {Munari}, {Nguyen},
  {Nordlander}, {Ocvirk}, {O'Flaherty}, {Olias Sanz}, {Ortiz}, {Osorio},
  {Oszkiewicz}, {Ouzounis}, {Palmer}, {Park}, {Pasquato}, {Peltzer}, {Peralta},
  {P{\'e}turaud}, {Pieniluoma}, {Pigozzi}, {Poels}, {Prat}, {Prod'homme},
  {Raison}, {Rebordao}, {Risquez}, {Rocca-Volmerange}, {Rosen}, {Ruiz-Fuertes},
  {Russo}, {Sembay}, {Serraller Vizcaino}, {Short}, {Siebert}, {Silva},
  {Sinachopoulos}, {Slezak}, {Soffel}, {Sosnowska}, {Strai{\v{z}}ys}, {ter
  Linden}, {Terrell}, {Theil}, {Tiede}, {Troisi}, {Tsalmantza}, {Tur},
  {Vaccari}, {Vachier}, {Valles}, {Van Hamme}, {Veltz}, {Virtanen}, {Wallut},
  {Wichmann}, {Wilkinson}, {Ziaeepour}, \& {Zschocke}}]{2016A&A...595A...1G}
{Gaia Collaboration}, {Prusti}, T., {de Bruijne}, J.~H.~J., {et~al.} 2016,
  \aap, 595, A1

\bibitem[{{Gregory} \& {Loredo}(1996)}]{1996ApJ...473.1059G}
{Gregory}, P.~C. \& {Loredo}, T.~J. 1996, \apj, 473, 1059

\bibitem[{{Grundstrom} \& {Gies}(2006)}]{2006ApJ...651L..53G}
{Grundstrom}, E.~D. \& {Gies}, D.~R. 2006, \apjl, 651, L53

\bibitem[{{Haberl} {et~al.}(2008){Haberl}, {Eger}, \&
  {Pietsch}}]{2008A&A...489..327H}
{Haberl}, F., {Eger}, P., \& {Pietsch}, W. 2008, \aap, 489, 327

\bibitem[{{Haberl} {et~al.}(2022{\natexlab{a}}){Haberl}, {Maitra}, {Carpano},
  {Dai}, {Doroshenko}, {Dennerl}, {Freyberg}, {Sasaki}, {Udalski}, {Postnov},
  \& {Shakura}}]{2022A&A...661A..25H}
{Haberl}, F., {Maitra}, C., {Carpano}, S., {et~al.} 2022{\natexlab{a}}, \aap,
  661, A25

\bibitem[{{Haberl} {et~al.}(2020){Haberl}, {Maitra}, {Carpano}, {Ducci},
  {Doroshenko}, {Koenig}, {Buckley}, {Monageng}, \&
  {Udalski}}]{2020ATel13609....1H}
{Haberl}, F., {Maitra}, C., {Carpano}, S., {et~al.} 2020, The Astronomer's
  Telegram, 13609, 1

\bibitem[{{Haberl} {et~al.}(2022{\natexlab{b}}){Haberl}, {Maitra},
  {Vasilopoulos}, {Maggi}, {Udalski}, {Monageng}, \&
  {Buckley}}]{2022A&A...662A..22H}
{Haberl}, F., {Maitra}, C., {Vasilopoulos}, G., {et~al.} 2022{\natexlab{b}},
  \aap, 662, A22

\bibitem[{{Haberl} {et~al.}(2021){Haberl}, {Salganik}, {Maitra}, {Doroshenko},
  {Ducci}, {Kaltenbrunner}, {Kreykenbohm}, {Lutovinov}, {Maggi}, {Mereminskiy},
  {Molkov}, {Rau}, {Semena}, {Tsygankov}, {Vasilopoulos}, {Weber}, \&
  {Wilms}}]{2021ATel15133....1H}
{Haberl}, F., {Salganik}, A., {Maitra}, C., {et~al.} 2021, The Astronomer's
  Telegram, 15133, 1

\bibitem[{{Haberl} \& {Sturm}(2016)}]{2016A&A...586A..81H}
{Haberl}, F. \& {Sturm}, R. 2016, \aap, 586, A81

\bibitem[{{Klus} {et~al.}(2013){Klus}, {Bartlett}, {Bird}, {Coe}, {Corbet}, \&
  {Udalski}}]{2013MNRAS.428.3607K}
{Klus}, H., {Bartlett}, E.~S., {Bird}, A.~J., {et~al.} 2013, \mnras, 428, 3607

\bibitem[{{Knigge} {et~al.}(2011){Knigge}, {Coe}, \&
  {Podsiadlowski}}]{2011Natur.479..372K}
{Knigge}, C., {Coe}, M.~J., \& {Podsiadlowski}, P. 2011, \nat, 479, 372

\bibitem[{{K{\"o}nig} {et~al.}(2022){K{\"o}nig}, {Saxton}, {Kretschmar},
  {Angelini}, {Belanger}, {Evans}, {Freyberg}, {Savchenko}, {Traulsen}, \&
  {Wilms}}]{2022A&C....3800529K}
{K{\"o}nig}, O., {Saxton}, R.~D., {Kretschmar}, P., {et~al.} 2022, Astronomy
  and Computing, 38, 100529

\bibitem[{{La Palombara} {et~al.}(2013){La Palombara}, {Mereghetti}, {Sidoli},
  {Tiengo}, \& {Esposito}}]{2013MmSAI..84..626L}
{La Palombara}, N., {Mereghetti}, S., {Sidoli}, L., {Tiengo}, A., \&
  {Esposito}, P. 2013, \memsai, 84, 626

\bibitem[{{Lomb}(1976)}]{1976Ap&SS..39..447L}
{Lomb}, N.~R. 1976, \apss, 39, 447

\bibitem[{{Long} {et~al.}(1981){Long}, {Helfand}, \&
  {Grabelsky}}]{1981ApJ...248..925L}
{Long}, K.~S., {Helfand}, D.~J., \& {Grabelsky}, D.~A. 1981, \apj, 248, 925

\bibitem[{{Luck} {et~al.}(1998){Luck}, {Moffett}, {Barnes}, \&
  {Gieren}}]{1998AJ....115..605L}
{Luck}, R.~E., {Moffett}, T.~J., {Barnes}, Thomas~G., I., \& {Gieren}, W.~P.
  1998, \aj, 115, 605

\bibitem[{{Maitra} {et~al.}(2020{\natexlab{a}}){Maitra}, {Haberl}, {Carpano},
  {Koenig}, {Doroshenko}, {Ducci}, {Buckley}, {Monageng}, \&
  {Udalski}}]{2020ATel13610....1M}
{Maitra}, C., {Haberl}, F., {Carpano}, S., {et~al.} 2020{\natexlab{a}}, The
  Astronomer's Telegram, 13610, 1

\bibitem[{{Maitra} {et~al.}(2020{\natexlab{b}}){Maitra}, {Haberl}, {Koenig},
  {Doroshenko}, {Carpano}, \& {Ducci}}]{2020ATel13650....1M}
{Maitra}, C., {Haberl}, F., {Koenig}, O., {et~al.} 2020{\natexlab{b}}, The
  Astronomer's Telegram, 13650, 1

\bibitem[{{Maitra} {et~al.}(2021){Maitra}, {Haberl}, {Vasilopoulos}, {Ducci},
  {Dennerl}, \& {Carpano}}]{2021A&A...647A...8M}
{Maitra}, C., {Haberl}, F., {Vasilopoulos}, G., {et~al.} 2021, \aap, 647, A8

\bibitem[{{Maitra} {et~al.}(2023){Maitra}, {Kaltenbrunner}, {Haberl},
  {Buckley}, {Monageng}, {Udalski}, {Carpano}, {Coley}, {Doroshenko}, {Ducci},
  {Malacaria}, {K{\"o}nig}, {Santangelo}, {Vasilopoulos}, \&
  {Wilms}}]{2023A&A...669A..30M}
{Maitra}, C., {Kaltenbrunner}, D., {Haberl}, F., {et~al.} 2023, \aap, 669, A30

\bibitem[{{Martin} {et~al.}(2011){Martin}, {Pringle}, {Tout}, \&
  {Lubow}}]{2011MNRAS.416.2827M}
{Martin}, R.~G., {Pringle}, J.~E., {Tout}, C.~A., \& {Lubow}, S.~H. 2011,
  \mnras, 416, 2827

\bibitem[{{Massey}(2002)}]{2002ApJS..141...81M}
{Massey}, P. 2002, \apjs, 141, 81

\bibitem[{{Meaburn}(1980)}]{1980MNRAS.192..365M}
{Meaburn}, J. 1980, \mnras, 192, 365

\bibitem[{{Monageng} {et~al.}(2020){Monageng}, {Coe}, {Buckley}, {McBride},
  {Kennea}, {Udalski}, {Evans}, {Clark}, \& {Negueruela}}]{2020MNRAS.496.3615M}
{Monageng}, I.~M., {Coe}, M.~J., {Buckley}, D.~A.~H., {et~al.} 2020, \mnras,
  496, 3615

\bibitem[{{Negueruela} \& {Coe}(2002)}]{2002A&A...385..517N}
{Negueruela}, I. \& {Coe}, M.~J. 2002, \aap, 385, 517

\bibitem[{{Okazaki}(2012)}]{2012ASPC..464..177O}
{Okazaki}, A.~T. 2012, in Astronomical Society of the Pacific Conference
  Series, Vol. 464, Circumstellar Dynamics at High Resolution, ed. A.~C.
  {Carciofi} \& T.~{Rivinius}, 177

\bibitem[{{Pavlinsky} {et~al.}(2021){Pavlinsky}, {Tkachenko}, {Levin},
  {Alexandrovich}, {Arefiev}, {Babyshkin}, {Batanov}, {Bodnar}, {Bogomolov},
  {Bubnov}, {Buntov}, {Burenin}, {Chelovekov}, {Chen}, {Drozdova}, {Ehlert},
  {Filippova}, {Frolov}, {Gamkov}, {Garanin}, {Garin}, {Glushenko}, {Gorelov},
  {Grebenev}, {Grigorovich}, {Gureev}, {Gurova}, {Ilkaev}, {Katasonov},
  {Krivchenko}, {Krivonos}, {Korotkov}, {Kudelin}, {Kuznetsova}, {Lazarchuk},
  {Lomakin}, {Lapshov}, {Lipilin}, {Lutovinov}, {Mereminskiy}, {Molkov},
  {Nazarov}, {Oleinikov}, {Pikalov}, {Ramsey}, {Roiz}, {Rotin}, {Ryadov},
  {Sankin}, {Sazonov}, {Sedov}, {Semena}, {Semena}, {Serbinov}, {Shirshakov},
  {Shtykovsky}, {Shvetsov}, {Sunyaev}, {Swartz}, {Tambov}, {Voron}, \&
  {Yaskovich}}]{2021A&A...650A..42P}
{Pavlinsky}, M., {Tkachenko}, A., {Levin}, V., {et~al.} 2021, \aap, 650, A42

\bibitem[{{Predehl} {et~al.}(2021){Predehl}, {Andritschke}, {Arefiev},
  {Babyshkin}, {Batanov}, {Becker}, {B{\"o}hringer}, {Bogomolov}, {Boller},
  {Borm}, {Bornemann}, {Br{\"a}uninger}, {Br{\"u}ggen}, {Brunner}, {Brusa},
  {Bulbul}, {Buntov}, {Burwitz}, {Burkert}, {Clerc}, {Churazov}, {Coutinho},
  {Dauser}, {Dennerl}, {Doroshenko}, {Eder}, {Emberger}, {Eraerds},
  {Finoguenov}, {Freyberg}, {Friedrich}, {Friedrich}, {F{\"u}rmetz},
  {Georgakakis}, {Gilfanov}, {Granato}, {Grossberger}, {Gueguen}, {Gureev},
  {Haberl}, {H{\"a}lker}, {Hartner}, {Hasinger}, {Huber}, {Ji}, {Kienlin},
  {Kink}, {Korotkov}, {Kreykenbohm}, {Lamer}, {Lomakin}, {Lapshov}, {Liu},
  {Maitra}, {Meidinger}, {Menz}, {Merloni}, {Mernik}, {Mican}, {Mohr},
  {M{\"u}ller}, {Nandra}, {Nazarov}, {Pacaud}, {Pavlinsky}, {Perinati},
  {Pfeffermann}, {Pietschner}, {Ramos-Ceja}, {Rau}, {Reiffers}, {Reiprich},
  {Robrade}, {Salvato}, {Sanders}, {Santangelo}, {Sasaki}, {Scheuerle},
  {Schmid}, {Schmitt}, {Schwope}, {Shirshakov}, {Steinmetz}, {Stewart},
  {Str{\"u}der}, {Sunyaev}, {Tenzer}, {Tiedemann}, {Tr{\"u}mper}, {Voron},
  {Weber}, {Wilms}, \& {Yaroshenko}}]{2021A&A...647A...1P}
{Predehl}, P., {Andritschke}, R., {Arefiev}, V., {et~al.} 2021, \aap, 647, A1

\bibitem[{{Reig} {et~al.}(1997){Reig}, {Fabregat}, \&
  {Coe}}]{1997A&A...322..193R}
{Reig}, P., {Fabregat}, J., \& {Coe}, M.~J. 1997, \aap, 322, 193

\bibitem[{{Riquelme} {et~al.}(2012){Riquelme}, {Torrej{\'o}n}, \&
  {Negueruela}}]{2012A&A...539A.114R}
{Riquelme}, M.~S., {Torrej{\'o}n}, J.~M., \& {Negueruela}, I. 2012, \aap, 539,
  A114

\bibitem[{{Rolleston} {et~al.}(2002){Rolleston}, {Trundle}, \&
  {Dufton}}]{2002A&A...396...53R}
{Rolleston}, W.~R.~J., {Trundle}, C., \& {Dufton}, P.~L. 2002, \aap, 396, 53

\bibitem[{{Salganik} {et~al.}(2022){Salganik}, {Tsygankov}, {Lutovinov},
  {Mushtukov}, {Mereminskiy}, {Molkov}, \& {Semena}}]{2022MNRAS.514.4018S}
{Salganik}, A., {Tsygankov}, S.~S., {Lutovinov}, A.~A., {et~al.} 2022, \mnras,
  514, 4018

\bibitem[{{Saxton} {et~al.}(2022){Saxton}, {K{\"o}nig}, {Descalzo}, {Belanger},
  {Kretschmar}, {Gabriel}, {Evans}, {Ibarra}, {Colomo}, {Sarmiento}, {Salgado},
  {Agrafojo}, \& {Kuulkers}}]{2022A&C....3800531S}
{Saxton}, R.~D., {K{\"o}nig}, O., {Descalzo}, M., {et~al.} 2022, Astronomy and
  Computing, 38, 100531

\bibitem[{{Scargle}(1982)}]{1982ApJ...263..835S}
{Scargle}, J.~D. 1982, \apj, 263, 835

\bibitem[{{Schmidtke} {et~al.}(1994){Schmidtke}, {Cowley}, {Frattare},
  {McGrath}, {Hutchings}, \& {Crampton}}]{1994PASP..106..843S}
{Schmidtke}, P.~C., {Cowley}, A.~P., {Frattare}, L.~M., {et~al.} 1994, \pasp,
  106, 843

\bibitem[{{Skrutskie} {et~al.}(2006){Skrutskie}, {Cutri}, {Stiening},
  {Weinberg}, {Schneider}, {Carpenter}, {Beichman}, {Capps}, {Chester},
  {Elias}, {Huchra}, {Liebert}, {Lonsdale}, {Monet}, {Price}, {Seitzer},
  {Jarrett}, {Kirkpatrick}, {Gizis}, {Howard}, {Evans}, {Fowler}, {Fullmer},
  {Hurt}, {Light}, {Kopan}, {Marsh}, {McCallon}, {Tam}, {Van Dyk}, \&
  {Wheelock}}]{2006AJ....131.1163S}
{Skrutskie}, M.~F., {Cutri}, R.~M., {Stiening}, R., {et~al.} 2006, \aj, 131,
  1163

\bibitem[{{Str{\"u}der} {et~al.}(2001){Str{\"u}der}, {Briel}, {Dennerl},
  {Hartmann}, {Kendziorra}, {Meidinger}, {Pfeffermann}, {Reppin}, {Aschenbach},
  {Bornemann}, {Br{\"a}uninger}, {Burkert}, {Elender}, {Freyberg}, {Haberl},
  {Hartner}, {Heuschmann}, {Hippmann}, {Kastelic}, {Kemmer}, {Kettenring},
  {Kink}, {Krause}, {M{\"u}ller}, {Oppitz}, {Pietsch}, {Popp}, {Predehl},
  {Read}, {Stephan}, {St{\"o}tter}, {Tr{\"u}mper}, {Holl}, {Kemmer}, {Soltau},
  {St{\"o}tter}, {Weber}, {Weichert}, {von Zanthier}, {Carathanassis}, {Lutz},
  {Richter}, {Solc}, {B{\"o}ttcher}, {Kuster}, {Staubert}, {Abbey}, {Holland},
  {Turner}, {Balasini}, {Bignami}, {La Palombara}, {Villa}, {Buttler},
  {Gianini}, {Lain{\'e}}, {Lumb}, \& {Dhez}}]{2001A&A...365L..18S}
{Str{\"u}der}, L., {Briel}, U., {Dennerl}, K., {et~al.} 2001, \aap, 365, L18

\bibitem[{{Sturm} {et~al.}(2013){Sturm}, {Haberl}, {Pietsch}, {Ballet},
  {Hatzidimitriou}, {Buckley}, {Coe}, {Ehle}, {Filipovi{\'c}}, {La Palombara},
  \& {Tiengo}}]{2013A&A...558A...3S}
{Sturm}, R., {Haberl}, F., {Pietsch}, W., {et~al.} 2013, \aap, 558, A3

\bibitem[{{Treiber} {et~al.}(2021){Treiber}, {Vasilopoulos}, {Bailyn},
  {Haberl}, {Gendreau}, {Ray}, {Maitra}, {Maggi}, {Jaisawal}, {Udalski},
  {Wilms}, {Monageng}, {Buckley}, {K{\"o}nig}, \&
  {Carpano}}]{2021MNRAS.503.6187T}
{Treiber}, H., {Vasilopoulos}, G., {Bailyn}, C.~D., {et~al.} 2021, \mnras, 503,
  6187

\bibitem[{{Turner} {et~al.}(2001){Turner}, {Abbey}, {Arnaud}, {Balasini},
  {Barbera}, {Belsole}, {Bennie}, {Bernard}, {Bignami}, {Boer}, {Briel},
  {Butler}, {Cara}, {Chabaud}, {Cole}, {Collura}, {Conte}, {Cros}, {Denby},
  {Dhez}, {Di Coco}, {Dowson}, {Ferrando}, {Ghizzardi}, {Gianotti}, {Goodall},
  {Gretton}, {Griffiths}, {Hainaut}, {Hochedez}, {Holland}, {Jourdain},
  {Kendziorra}, {Lagostina}, {Laine}, {La Palombara}, {Lortholary}, {Lumb},
  {Marty}, {Molendi}, {Pigot}, {Poindron}, {Pounds}, {Reeves}, {Reppin},
  {Rothenflug}, {Salvetat}, {Sauvageot}, {Schmitt}, {Sembay}, {Short},
  {Spragg}, {Stephen}, {Str{\"u}der}, {Tiengo}, {Trifoglio}, {Tr{\"u}mper},
  {Vercellone}, {Vigroux}, {Villa}, {Ward}, {Whitehead}, \&
  {Zonca}}]{2001A&A...365L..27T}
{Turner}, M.~J.~L., {Abbey}, A., {Arnaud}, M., {et~al.} 2001, \aap, 365, L27

\bibitem[{{Udalski} {et~al.}(2008){Udalski}, {Szymanski}, {Soszynski}, \&
  {Poleski}}]{2008AcA....58...69U}
{Udalski}, A., {Szymanski}, M.~K., {Soszynski}, I., \& {Poleski}, R. 2008,
  \actaa, 58, 69

\bibitem[{{Udalski} {et~al.}(2015){Udalski}, {Szyma{\'n}ski}, \&
  {Szyma{\'n}ski}}]{2015AcA....65....1U}
{Udalski}, A., {Szyma{\'n}ski}, M.~K., \& {Szyma{\'n}ski}, G. 2015, \actaa, 65,
  1

\bibitem[{{Vasilopoulos} {et~al.}(2013){Vasilopoulos}, {Maggi}, {Haberl},
  {Sturm}, {Pietsch}, {Bartlett}, \& {Coe}}]{2013A&A...558A..74V}
{Vasilopoulos}, G., {Maggi}, P., {Haberl}, F., {et~al.} 2013, \aap, 558, A74

\bibitem[{{Vasilopoulos} {et~al.}(2017){Vasilopoulos}, {Zezas}, {Antoniou}, \&
  {Haberl}}]{2017MNRAS.470.4354V}
{Vasilopoulos}, G., {Zezas}, A., {Antoniou}, V., \& {Haberl}, F. 2017, \mnras,
  470, 4354

\bibitem[{{Verner} {et~al.}(1996){Verner}, {Ferland}, {Korista}, \&
  {Yakovlev}}]{1996ApJ...465..487V}
{Verner}, D.~A., {Ferland}, G.~J., {Korista}, K.~T., \& {Yakovlev}, D.~G. 1996,
  \apj, 465, 487

\bibitem[{{Wilms} {et~al.}(2000){Wilms}, {Allen}, \&
  {McCray}}]{2000ApJ...542..914W}
{Wilms}, J., {Allen}, A., \& {McCray}, R. 2000, \apj, 542, 914

\bibitem[{{Zaritsky} {et~al.}(2004){Zaritsky}, {Harris}, {Thompson}, \&
  {Grebel}}]{2004AJ....128.1606Z}
{Zaritsky}, D., {Harris}, J., {Thompson}, I.~B., \& {Grebel}, E.~K. 2004, \aj,
  128, 1606

\end{thebibliography}

\begin{appendix}
\section{\ero light curves}
\begin{figure*}[h]
 \resizebox{\hsize}{!}{\includegraphics{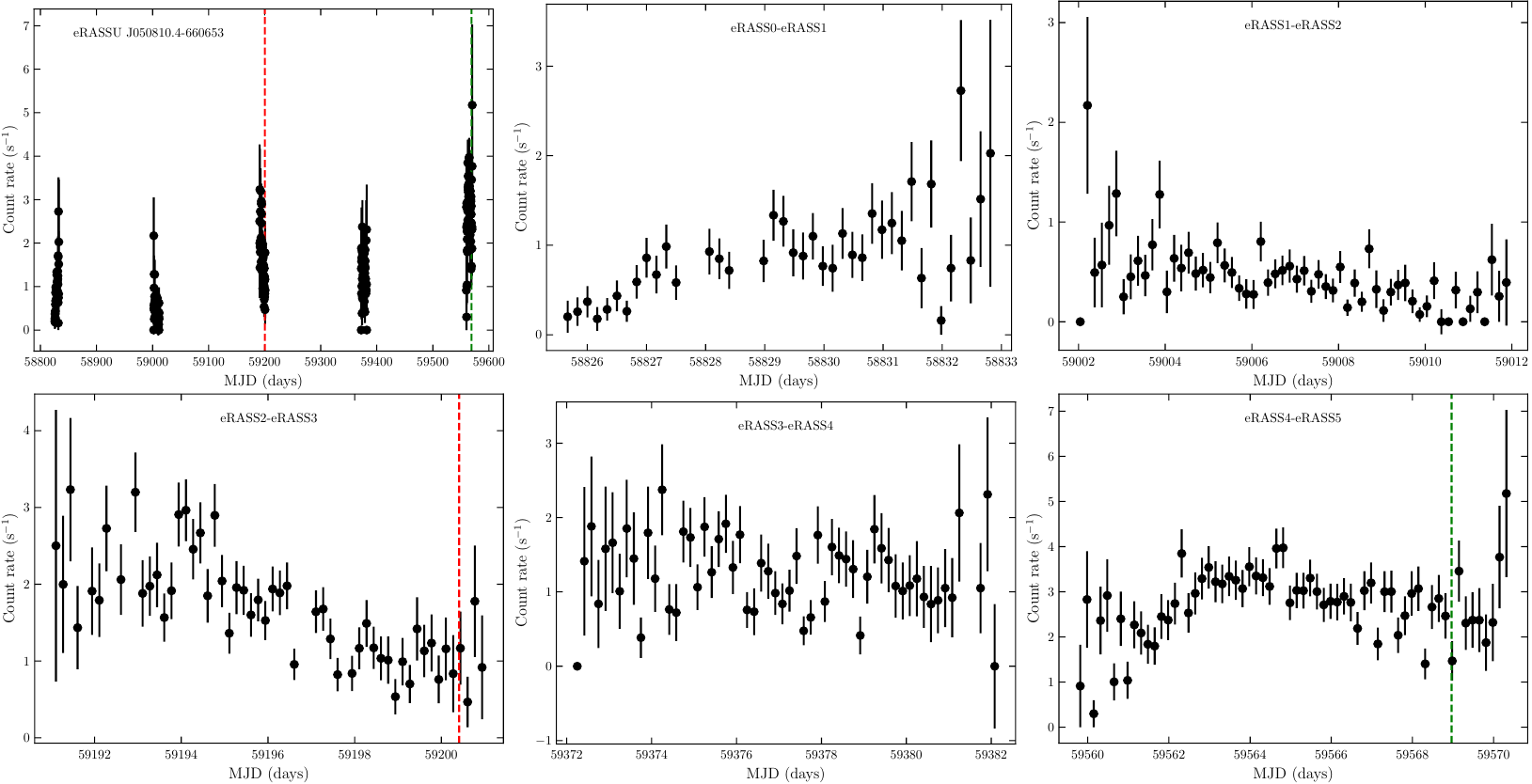}}
  \caption{
    \ero light curve of \asrc in the energy band 0.2--8\,keV, combining data from all TMs. Until the end of 2021, \ero scanned the source during five epochs (upper left). The first scans were performed during a test phase (named eRASS0) before the official start of the all-sky survey (MJD 58828) with eRASS1 and then around the formal transitions of eRASS1-eRASS2, eRASS2-eRASS3, eRASS3-eRASS4, and eRASS4-eRASS5. The red and green dashed lines mark the beginning of the \xmm and \nustar observations \citep{2022MNRAS.514.4018S,2021ATel15133....1H}, respectively.
  }
  \label{fig:erolca}
\end{figure*}
\begin{figure*}
 \resizebox{\hsize}{!}{\includegraphics{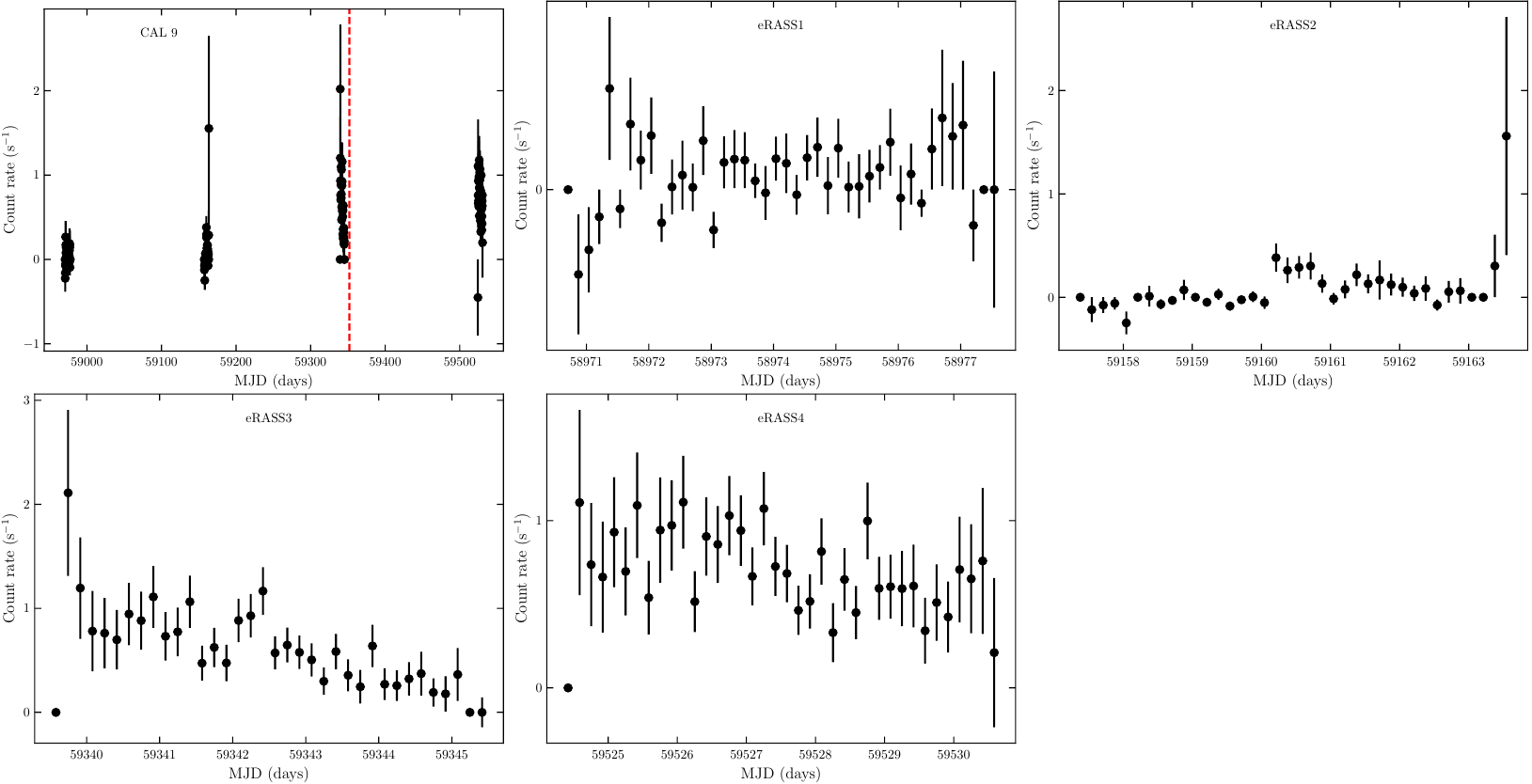}}
  \caption{
     \ero light curve of \bsrc, created as described in Fig.\,\ref{fig:erolca}.
     \ero scanned the source during eRASS1--4. Red dashed line marks the beginning of the \xmm observation, which was performed shortly after the last source scans during eRASS3.
  }
  \label{fig:erolcb}
\end{figure*}
\begin{figure*}
 \resizebox{0.666\hsize}{!}{\includegraphics{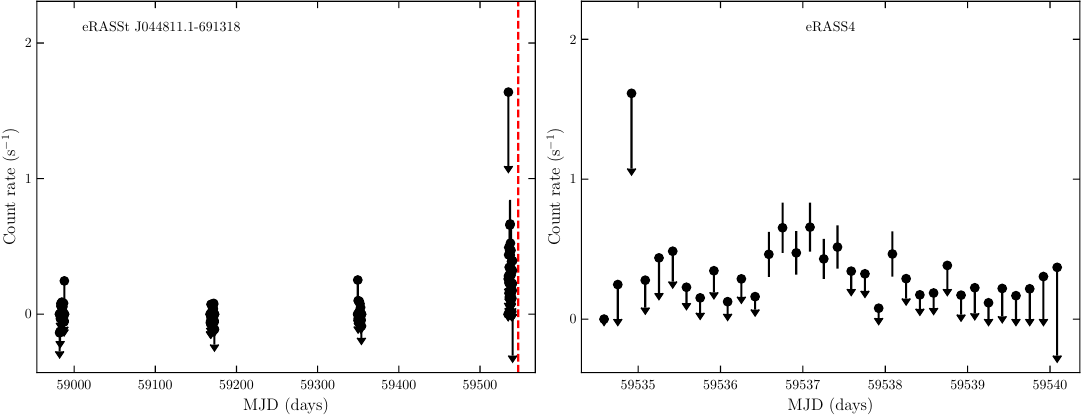}}
  \caption{
    \ero light curve of \csrc, created as described in Fig.\,\ref{fig:erolca}.
    The red dashed line marks the beginning of the \xmm observation, which was performed shortly after the last source scans during eRASS4. The source was not detected during eRASS1, 2, and 3 (the corresponding light curves are not shown).
  }
  \label{fig:erolcc}
\end{figure*}

\end{appendix}
\end{document}